\documentclass[letterpaper,twocolumn,english,preprintnumbers,amsmath,amssymb,superscriptaddress,footinbib,accepted=2026-07-15]{quantumarticle}
\pdfoutput=1
\usepackage{tikz}
\usepackage[many]{tcolorbox}
\usepackage[numbers]{natbib}
\usetikzlibrary{shapes.geometric, arrows}
\usetikzlibrary{calc}

\tikzstyle{startstop} = [rectangle, rounded corners, 
minimum width=3cm, 
minimum height=1cm,
text centered, draw=black]

\tikzstyle{io} = [trapezium, 
trapezium stretches=true, % A later addition
trapezium left angle=70, 
trapezium right angle=110, 
minimum width=3cm, 
minimum height=1cm, text centered, 
draw=black]

\tikzstyle{process} = [rectangle, 
minimum width=3cm, 
minimum height=1cm, 
text centered, 
text width=3cm, 
draw=black]

\tikzstyle{decision} = [diamond, 
minimum width=3cm, 
minimum height=1cm, 
text centered, 
draw=black]

\tikzstyle{arrow} = [thick,->,>=stealth]
\usepackage{graphicx}

\usepackage{catchfile}

\usepackage{amsthm}
\usepackage{amssymb}
\usepackage{mathtools}
\usepackage{nicefrac}
\usepackage{fontawesome}

\graphicspath{{./Images/}}
\usepackage{xcolor}
\usepackage[colorlinks = true, citecolor=blue, linkcolor=black,urlcolor=purple]{hyperref}
\usepackage{enumitem}
    \setlist[itemize]{noitemsep, topsep=0pt}
    \setlist[enumerate]{noitemsep, topsep=0pt}
\usepackage{verbatim}
\usepackage{mathtools,amssymb,amsmath}
\usepackage{bm}
\usepackage{cancel}
\usepackage{relsize}
\usepackage{array}
\usepackage{bm}
\usepackage{subfigure}

\newcommand{\ee}{\end{equation}}

\usepackage[normalem]{ulem}
\usepackage{color}

\usepackage{afterpage}

\newcommand{\jila}{JILA, University of Colorado and National Institute of Standards and Technology,
and Department of Physics, University of Colorado, Boulder, CO 80309, USA}
\newcommand{\pme}{Pritzker School of Molecular Engineering, University of Chicago, Chicago, IL 60637, USA}
\newcommand{\ntu}{Department of Physics and Department of Computer Science and Information Engineering, National Taiwan University, Taipei 10617, Taiwan}

\newcommand{\harvardqse}{Quantum Science and Engineering, Harvard Griffin Graduate School of Arts and Sciences, Harvard University, Cambridge, MA 02138, USA}
\newcommand{\psd}{Department of Physics, University of Chicago, Chicago, IL 60637, USA}
\newcommand{\iqoqi}{Institute for Quantum Optics and Quantum Information,
Austrian Academy of Sciences, 6020 Innsbruck, Austria}
\newcommand{\uibk}{Institute for Experimental Physics, University of Innsbruck, 6020 Innsbruck, Austria}

\usepackage[ruled, lined, resetcount, longend]{algorithm2e}
\SetAlgoCaptionLayout{plain}
\SetNlSty{}{}{}

\tcolorboxenvironment{problem}{
  colback=black!10,
  colframe=white,%
  width=\dimexpr\linewidth+10pt\relax,%  
  enlarge left by=-5pt,%                  
  enlarge right by=-5pt,%               
  breakable,
  boxsep=5pt,%            
  boxrule=0pt,
  left=0pt,right=0pt,top=0pt,bottom=0pt,
  sharp corners,
  before skip=\topsep,
  after skip=\topsep
}

\newenvironment{problem*}
  {\begin{tcolorbox}[colback=black!10,
                     colframe=white,
                     width=\dimexpr\linewidth+10pt\relax,
                     enlarge left by=-5pt,
                     enlarge right by=-5pt,
                     breakable,
                     boxsep=5pt,
                     boxrule=0pt,
                     left=0pt,right=0pt,top=0pt,bottom=0pt,
                     sharp corners,
                     before skip=\topsep,
                     after skip=\topsep]}
  {\end{tcolorbox}}

\theoremstyle{definition}

\tcolorboxenvironment{definition}{
  colback=black!10,
  colframe=white,
  width=\dimexpr\linewidth+10pt\relax,
  enlarge left by=-5pt,
  enlarge right by=-5pt,
  boxsep=5pt,
  boxrule=0pt,
  left=0pt,right=0pt,top=0pt,bottom=0pt,
  sharp corners,
  before skip=\topsep,
  after skip=\topsep
}

\theoremstyle{remark}

\tcolorboxenvironment{remark}{
  colback=black!10,
  colframe=white,
  width=\dimexpr\linewidth+10pt\relax,
  enlarge left by=-5pt,
  enlarge right by=-5pt,
  boxsep=5pt,
  boxrule=0pt,
  left=0pt,right=0pt,top=0pt,bottom=0pt,
  sharp corners,
  before skip=\topsep,
  after skip=\topsep
}
  
\makeatletter
\newsavebox{\@brx}
\newcommand{\llangle}[1][]{\savebox{\@brx}{\(\m@th{#1\langle}\)}
  \mathopen{\copy\@brx\kern-0.5\wd\@brx\usebox{\@brx}}}
\newcommand{\rrangle}[1][]{\savebox{\@brx}{\(\m@th{#1\rangle}\)}
  \mathclose{\copy\@brx\kern-0.5\wd\@brx\usebox{\@brx}}}
\makeatother

\usepackage{xcolor}

\makeatletter 
\renewcommand{\fnum@figure}{\textbf{Fig.~\thefigure}}
\def\@caption@fignum@sep{\textbf{.} }
\makeatother 

\makeatletter
\appto{\@maketitle}{%
  \rmfamily\normalsize
  \vskip 0.25em
  \begin{quotation}
The task of atom rearrangement has emerged in the last decade as a fundamental building block in the development of neutral atom-based quantum processors. 
As such processors grow to thousands of atoms, it becomes increasingly important to design algorithms robust to experimental sources of error. While recent progress has been made towards developing algorithms with favorable time scaling, such work has been limited to noiseless settings. Moreover, there is a lack of open-source code for reproducing and benchmarking existing algorithms.
To address these deficiencies, we develop an open-source simulation framework, \textit{atommovr}, and leverage it to study three distinct settings: 1) time-optimal, noiseless rearrangement, 2) noisy rearrangement under realistic error models, and 3) noiseless dual-species rearrangement. We extract lower bounds for time-optimal rearrangement, study advantageous strategies across different error regimes, and develop a novel dual-species algorithm, InsideOut, capable of avoiding `blocked' configurations with a near-unity success rate. We hope that \textit{atommovr} can serve as a common tool for the community to study rearrangement, lower the barrier to entry for new experimental groups, and stimulate progress in developing algorithms tailored to minimize atom loss in experiment.

\begin{center}
\faGithub~\href{https://github.com/bernienlab/atommovr}{Install \textit{atommovr} here}
\end{center}
  \end{quotation}
  \vskip 0.5em
}
\makeatother

\begin{document}
    
\makeatother

\title{atommovr: An open-source simulation framework for rearrangement in atomic arrays}

\author{Nikhil K. Harle}
\affiliation{\pme}
\affiliation{\jila}
\listcsgadd{author1affiliations}{\ref{fn:equal}}

\author{Bo-Yu Chen}
\affiliation{\pme}
\affiliation{\iqoqi}
\affiliation{\ntu}
\listcsgadd{author2affiliations}{\ref{fn:equal}}

\author{Bob Bao}
\affiliation{\psd}
\affiliation{\harvardqse}

\author{Hannes Bernien}
\affiliation{\pme}
\affiliation{\iqoqi}
\affiliation{\uibk}

\maketitle

{\def\thefootnote{*}
\footnotetext{%
  \textsf{\label{fn:equal}These authors contributed equally to this work.}%
}}

\vspace{0.15cm}

\section{Background}

In recent years, the field of quantum computing has welcomed a new platform: quantum processors comprised of individual atoms trapped in optical tweezers~\cite{Lee2016, Barredo2016, Endres2016, Kim2016, Graham2022, Bluvstein2022}. 
Atomic quantum processors are versatile and flexible platforms whose utility ranges from exploring fundamental physics to implementing early paradigms of quantum computing \cite{ Reichardt2024, Bluvstein2024, Zhang2025}.

\begin{figure*}[t!]
\setlength{\tabcolsep}{10pt}
\centering
\includegraphics[width=\textwidth]
{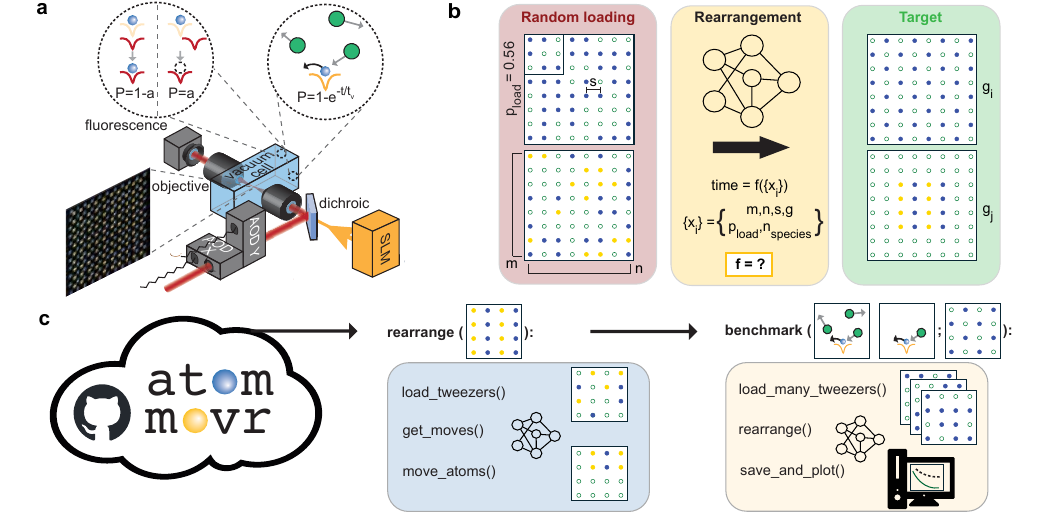}
\caption{\textbf{Schematic depiction of rearrangement}. 
\textbf{a}, 
An example experimental setup. 
Moving tweezers are generated by a pair of crossed AODs (gray boxes marked ``AOD X" and ``AOD Y"), which transform a single input tweezer into a rectilinear grid of tweezers, controlled by radio frequency (RF) inputs (black curves). 
Static tweezers are generated by an SLM (orange box). 
There are two predominant sources of error: loss due to collisions with background gas particles (large green dots) and tweezer losses due to transport and transfers between static and moving tweezers (orange and red curves).
\textbf{b}, Theoretical formulation of the time-optimal rearrangement problem.
Given a randomly loaded initial configuration $\mathbf{X_0}$, the task is to transform it into a specified target configuration $\mathbf{g}$ in the shortest possible time.
\textbf{c}, Overview of our open-source rearrangement simulation framework and pseudocode demonstrating its main functions:
simulating the rearrangement process and benchmarking under customizable error and timing models.
All parameters and error processes depicted in \textbf{a} and \textbf{b} are implemented in the simulation framework.}
\label{fig1}
\end{figure*}

A key subprocess of neutral atom processors is the physical rearrangement of optical tweezers to move atoms into particular geometric configurations (Fig.~\ref{fig1}b). 
% The loading of atoms into tweezers is a probabilistic process, and as such, rearrangement is universally necessary to prepare defect-free geometries. 
The flexibility of target geometry is a key advantage of atomic quantum processors over solid-state quantum processors, such as superconducting qubits. 
This ability, coupled with the ability to rearrange atoms in the middle of an experimental cycle, opens many possibilities, such as realizing many-body observables in quantum simulation and shuttling atoms between zones in various error correction schemes~\cite{Bluvstein2022, Xu2024, Viszlai2024, Poole2024}.
In addition, rearrangement can be leveraged to realize effective cooling by removing entropy \cite{B&Ctheory, shaw2025erasure, kumar2018sorting}.

\begin{figure*}[t!]
\setlength{\tabcolsep}{10pt}
\centering
\includegraphics[]
{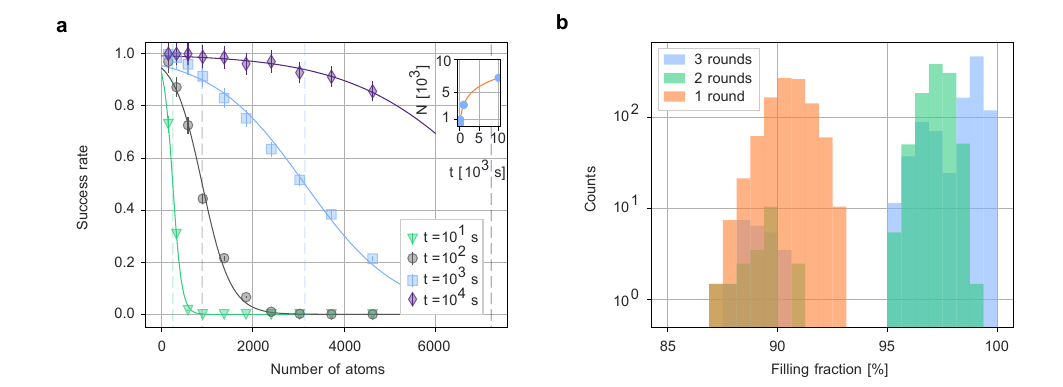}
\caption{\textbf{Algorithm performance under finite vacuum-limited lifetime}. \textbf{a}, Impact of vacuum-limited lifetime $t_\text{v}$ on achievable array sizes for the Balance and Compact algorithm. 
Each data point is averaged over 500 randomly loaded initial configurations, and the error bars represent shot noise.
Solid lines are curve fits to a sigmoid function. For each curve, the number of atoms $n_c$ at which the success rate is 50\% is marked by a dashed vertical line (extracted from the curve fit).
Inset: extracted values of $n_c$ plotted as a function of the vacuum-limited lifetime. 
\textbf{b}, Mitigating the impact of vacuum-limited lifetime through consecutive rearrangement rounds for the Hungarian algorithm.
To mitigate atom loss, it can be effective to perform rearrangement multiple times, interspersed with images to check the final position of the atoms. We fix the vacuum-limited lifetime at $t_\text{v} = 15~\text{s}$, and the target is a fully filled $36\times36$ square centered in a $50\times50$ array.
Darker colors indicate regions of overlap between the histograms.}
\label{fig2}
\end{figure*}

As atomic quantum processors continue to mature and grow in size, so too does the need for efficient and scalable rearrangement algorithms: 
the probability of successfully generating a defect-free atom array decreases exponentially with atom number. While the vacuum-limited lifetime of a single atom is typically on the order of $10-100~\text{s}$ \cite{Schymik2020},
%, weighted by the vacuum-limited lifetime of a single atom, defined as $t_\text{v}$:
%\begin{equation}
%    P_\text{defect-free} = 1-\exp{(-Nt/t_{\text{v}})}
%\end{equation}
%Currently, typical vacuum-limited lifetimes are on the order of $10-100~\text{s}$ \cite{Schymik2020}, although recent experiments with optimized vacuum design have yielded times in excess of $1000~\text{s}$ \cite{Zhang2024, pichard2024rearrangement}.
the corresponding probability of maintaining a defect-free array of 1000 atoms falls to 50\% after only $7-70~\text{ms}$. Given that the time to move an atom by a single site already takes around $0.5~\text{ms}$ (including times for trap transfers), this imposes a severe practical constraint.
While recent experiments have realized lifetimes in excess of $1000~\text{s}$ with optimized vacuum design \cite{Zhang2024, pichard2024rearrangement}, the scaling is not favorable: a linear increase in vacuum-limited lifetime leads to a sublinear increase in the size of defect-free target configurations (Fig.~\ref{fig2}a). As multiple groups \cite{pichard2024rearrangement, Manetsch2024, Norcia2024, Pause24, Gyger2024} have now successfully prepared tweezer arrays with over 1000 atoms, there is a timely impetus to develop scalable methods.
% While continuous loading offers a path to large-scale quantum processor, the majority of current and near-term experiments rely on static loading, where algorithmic efficiency is the primary bottleneck against vacuum loss. 
% While continuous loading techniques have demonstrated large arrays \cite{Gyger2024, Norcia2024, chiu2025continuous}, standard static-loading architectures remain the primary platform for quantum information
% \cite{zhang2025fast, Pause24, pichard2024rearrangement, Manetsch2024}, where scaling is strictly limited by the ratio of rearrangement time to vacuum lifetime.

We note that there are opportunities for mitigation: a target may be incrementally constructed over many rearrangement rounds (Fig.~\ref{fig2}b), and continuous reloading techniques \cite{Norcia2024, Pause24, Gyger2024} may offer further mitigation by periodically replenishing atom numbers to compensate for losses. Nevertheless, scalable rearrangement is still desirable in this setting to reduce experimental cycle time and maximize the size of practically achievable configurations.

In addition to the limitations imposed by vacuum-limited lifetimes, another relevant concern is the impact of atom loss from transport and transfer between static and moving traps, which is commonly encountered in experimental settings (Fig.~\ref{fig1}a). However, these processes have not been thoroughly studied, let alone in conjunction with vacuum-limited lifetime. 

In this work, we introduce a simulation platform for benchmarking rearrangement algorithms under various error and timing models, and use it to look at the rearrangement problem from three different angles. In Section~\ref{sec:time-optimal-scaling}, we extract lower bounds for time-optimal rearrangement and extract scaling coefficients for various algorithms. In Section~\ref{sec:noisy}, we extend our benchmarking to noisy environments and observe the advantages of different heuristic strategies across error regimes. Lastly, in Section~\ref{sec:dual}, we examine the challenges of preparing target patterns comprised of two atomic species, and develop a novel algorithm, InsideOut, capable of circumventing these challenges.

\section{Definitions and existing work}
\textbf{Constraints on array and move structure}.
In this work, we restrict our focus to rearrangement on a square lattice (though extensions to Bravais lattices can be easily made) and operations $\epsilon$ performed by two crossed acousto-optical deflectors (AODs; see Fig.~\ref{fig1}a). While alternative move structures exist \cite{Kim2016, Treptow2021, Knottnerus2025, Wei2025} (see Section~\ref{sec:Discussion}), we do not actively study them in this work. 

The crossed-AOD device transforms a single incident beam into a movable rectilinear grid of beams that can be displaced, expanded, and/or contracted. 
This approach is extremely advantageous in that it allows for the simultaneous implementation of many single-atom displacements in a single `AOD move segment', which we define as an AOD operation which displaces one or more atoms by a single site in the horizontal and/or vertical directions (including diagonally). A multi-site move (i.e. where an atom is moved by $>1$ site) is described by a consecutive sequence of AOD move segments.

\textbf{Defining time-optimal rearrangement}. 
Consider a spatial configuration of $m \times n$ atoms, denoted as $\mathbf{X}\in\{0,1\}^{m\times n}$, where $\mathbf{X}_{ij} = 1~(0)$ indicates the presence (absence) of an atom in the tweezer located at the $i$th row and the $j$th column.
The rearrangement problem we consider is the following: given an initial configuration $\mathbf{X}_0$ and a final configuration $\mathbf{g}$, how many operations does it take to transform $\mathbf{X}_0 \xrightarrow{} \mathbf{g}$, and how does this scale with the total number of target atoms $N = \sum_{ij} \mathbf{g}_{ij} $ (Fig. \ref{fig1}b)?
More formally, the time-optimal rearrangement algorithm aims to address the following two questions:

\begin{problem*}
Let $\Omega$ be a finite configuration space and $S\subseteq \{\epsilon: \Omega \to \Omega\}$ a finite set of operators which each describe a single `AOD move segment'  (as defined in the previous section).

Given an initial atom array configuration $\mathbf{X_0}\in\Omega$ and a target configuration $\mathbf{g}\in\Omega$, what is $k^\star \coloneqq \min \{k\in\mathbb{N}~\vert~\exists~\epsilon_1,\ldots,\epsilon_k\in S~\text{ s.t. }~\epsilon_k\circ\cdots\circ\epsilon_1(\mathbf{X}_0)=\mathbf{g}\}$?

How does one explicitly construct 
a sequence $\epsilon_1,\ldots, \epsilon_{k^{\star}}$
achieving $\epsilon_{k^{\star}}\circ\cdots\circ\epsilon_1(\mathbf{X}_0)=\mathbf{g}$?
\end{problem*}

\textbf{Existing algorithms}.
There are two general approaches to designing algorithms for the rearrangement problem. 
The first is based on the `Hungarian' method of solving the linear sum assignment problem (LSAP) \cite{Schymik2020, Konig1916, Egervary1931, Kuhn1955, Lee2017}. Algorithms which follow this method typically utilize only a single moving tweezer, but are optimal with respect to \textit{transport distance}, as the solution to the LSAP minimizes the sum of the distances travelled by each individual atom. 

The second approach aims to minimize the total rearrangement time by exploiting the ability to move multiple atoms simultaneously with AODs. Such algorithms heavily utilize parallelized operations on single rows and columns, and a typical structure is to transfer atoms between rows such that each has a sufficient number of atoms, and subsequently condense each column to form a defect-free array. Representative algorithms include Parallel Sort and Compression \cite{Tian2023}, Parallel Compression Filling \cite{zhang2025fast}, Redistribution-Reconfiguration \cite{Cimring2023}, Balance and Compact \cite{B&C}, and Tetris \cite{Wang2023}.

\section{Rearrangement simulation environment}

We develop and demonstrate the use of a Python-based simulation framework, \emph{atommovr}, with the following features:
\begin{enumerate}
    \item \textbf{Benchmarking}. \textit{atommovr} enables easy comparison of the performance of various algorithms through automated calculation of relevant quantities such as rearrangement time, filling fraction, and success rate (Fig.~\ref{fig2}).

    \item\textbf{Library of existing algorithms}.
    \textit{atommovr} collects a variety of representative single-species and dual-species atom rearrangement algorithms, including the Hungarian~\cite{Schymik2020,Konig1916,Egervary1931,Kuhn1955,Lee2017}, an implementation of Balance and Compact~\cite{B&Ctheory,B&C} for AODs, and algorithms that we develop, namely ParHungarian (see \ref{sec:par_hung_single} for detailed descriptions) for both single-species and dual-species settings and InsideOut (see \ref{sec:inside_out} for detailed description) for the dual-species rearrangement.
    All algorithmic source code is publicly available on GitHub.
    
    \item \textbf{Realistic experimental conditions and timing parameters}. 
    We simulate the process of stochastic loading, atom loss due to vacuum-limited lifetime, and atom loss and move failures due to transfers between static and moving traps and acceleration/deceleration in moving tweezers.
    Furthermore, our simulations include customizable timing models (see \ref{sec:timing_models} for details) and a number of relevant experimental parameters, such as lattice spacing, maximum transport speed, additional time costs for trap transfers and acceleration/deceleration, and loading rate (Fig.~\ref{fig1}a).
    All of the features described above are easily user-programmable.

    \item \textbf{Visualization}. 
    \textit{atommovr} can generate movies of the rearrangement process to facilitate insight into algorithmic performance. 
    These include visualizations for various error processes, such as atom loss due to tweezer failure or collisions with background molecules.
    
    %\item \textbf{Customizable experimental toolkit}.
    %We intentionally developed \textit{atommovr} in such a way that it would be easily accessible and modifiable to serve various purposes, from finding a suitable algorithm for an experimental setup (Fig.~\ref{fig4}) to studying fundamental limitations. We have written tutorial notebooks that demonstrate the intended usage and walk readers through the process of building their own algorithms. In addition, the repository is modular and customizable to facilitate further feature development from the community.

\end{enumerate}

\section{Time-optimal scaling and benchmarks}
\label{sec:time-optimal-scaling}
\begin{figure}[t!]
\setlength{\tabcolsep}{10pt}
\centering
\includegraphics[width=\columnwidth]
{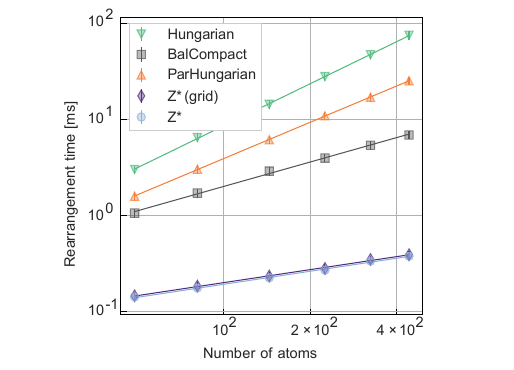}
\caption{\textbf{Algorithmic scaling and comparison to fundamental lower bounds in a noiseless setting}. 
Scaling behavior of two foundational rearrangement algorithms, Balance and Compact (gray squares), and the Hungarian (green triangles), as well as a variation of the Hungarian utilizing move parallelization (ParHungarian) (orange triangles). This behavior is contrasted with extracted lower bounds for the optimal rearrangement time for both unconstrained moves `$Z^\star$' (blue circles) and moves restricted to a grid `$Z^\star$' \textit{(grid)} (purple diamonds).
Each data point is averaged over 100 randomly loaded initial configurations, and the error bars represent shot noise. All target configurations are unit filling square grids. Scaling coefficients are presented in Table \ref{scaling}.}
\label{fig4}
\end{figure}

\textbf{Lower bounds for time-optimal scaling}.
We make use of \textit{atommovr} to numerically extract lower bounds for time-optimal rearrangement algorithms and compare the scaling behavior of algorithms recently proposed in the literature (Fig.~\ref{fig4}, Table~\ref{scaling}).
To extract lower bounds, we solve the \textit{linear bottleneck assignment problem} (LBAP) \cite{PPSU}, also known as the \textit{maximum bottleneck matching problem} (MBMP) \cite{Burkard1980} in the graph-theoretic literature, which closely resembles LSAP. 
The key difference is that instead of minimizing the \textit{total} distance traveled by all atoms, the solution to the LBAP/MBMP minimizes the maximum distance traveled by any particular atom. 
Assuming that the speed of a moving tweezer is fixed, a lower bound on the rearrangement time $t_{\text{LB}}$ can be easily extracted by the following formula:
\begin{equation}
    t_\text{LB} = \frac{s\times d_\text{minmax}}{\lvert v_\text{tweezer}\rvert} 
    %+ t_\text{accel} + t_\text{decel} + t_\text{pickup} + t_\text{putdown}
\end{equation}
where $s$ is the spacing between adjacent sites, $d_\text{minmax}$ is the solution to the LBAP/MBMP and $|v_\text{tweezer}|$ is the tweezer speed.
%the maximum speed reached by a tweezer, $t_\text{accel~(decel)}$ are additional time costs to account for nonlinear ramp profiles, and $t_\text{pickup~(putdown)}$ are additional time costs for transferring atoms from static (moving) to moving (static) tweezers.
We use the protocol and repository developed in~\cite{PPSU} to solve the LBAP/MBMP, which we refer to as `$Z^\star$'.

\textbf{Extracting scaling coefficients for heuristic algorithms}. Our benchmarking results are shown in Fig.~\ref{fig4} and we present extracted scaling coefficients in Table~\ref{scaling}.
Notably, for a 50\% loading probability, the lower bound scales as just under $\sqrt{N}$, while one of the leading algorithms (Tetris~\cite{Wang2023}) scales as just over $N$ (where $N$ is the number of target sites in a square grid with unit filling). While our implementation of the Balance and Compact for tweezer arrays is able to achieve sublinear scaling ($N^{0.84(2)}$), there is still a sizable gap.

Moreover, we develop a variation of the Hungarian algorithm with move parallelization (ParHungarian), and compare its scaling to our implementation of the naive Hungarian, which we expect to be roughly $\mathcal{O}(N\sqrt{N})$ (as this entails moving $\mathcal{O}(N)$ atoms sequentially, each with a maximum distance of $\sqrt{N}$).
We find that our extracted scaling agrees (within error bars) with that previously found in \cite{Wang2023}, and furthermore, observe that the naive addition of parallelization to the Hungarian algorithm improves the scaling (from $N^{1.45(1)}$ to $N^{1.26(1)}$).

\begin{table}[ht]
    \caption{Extracted scaling coefficients for rearrangement algorithms and theoretical lower bounds. All data is taken with a loading probability of 50\%.}
    \begin{center}
    \begin{tabular}{||c c c||} 
     \hline
     Algorithm & Scaling coefficient  & Source \\
        & $b~~(t(N) = cN^b)$ &  \\ [0.5ex] 
     \hline\hline
     $Z^\star$ bound & 0.45(6) & This work \\ 
     \hline
     
     $Z^\star$ (grid) bound & 0.45(6) & This work \\
     \hline

     BalanceCompact & 0.84(2) & This work \\ 
     \hline

     Tetris & 1.03(7) & \cite{Wang2023} \\ 
     \hline

     ParHungarian & 1.26(1) & This work \\
     \hline

     Hungarian & 1.45(1) & This work \\
     \hline

     Hungarian & 1.6(1) & \cite{Wang2023} \\ 
     \hline
    \end{tabular}
    \end{center}
    \label{scaling}
\end{table}

\section{Comparison of heuristic strategies under noise models}
\label{sec:noisy}
\begin{figure}[t!]
\setlength{\tabcolsep}{10pt}
\centering
\includegraphics[width=\columnwidth]
{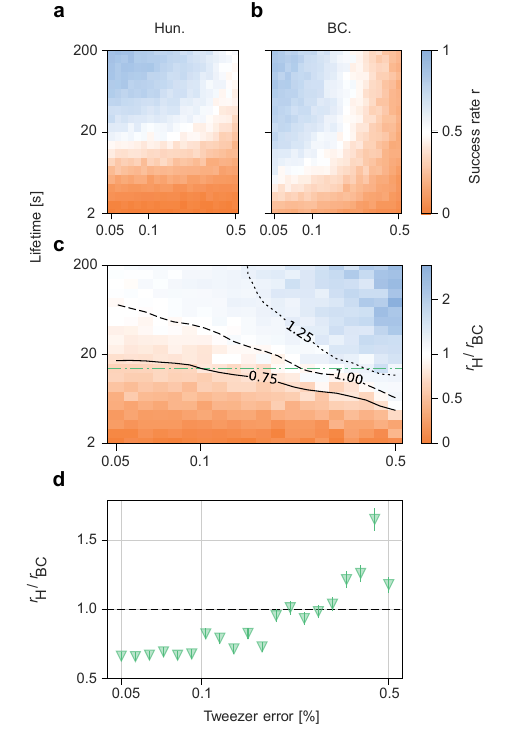}
\caption{\textbf{Comparing the choice of algorithms under realistic experimental conditions.} Rearrangement success rate after a single round for 
\textbf{a}, the Hungarian algorithm (which uses a single tweezer) and 
\textbf{b}, the Balance and Compact algorithm (which moves many atoms in parallel in sequential row-column operations), as a function of the vacuum-limited lifetime (y-axis) and the probability of losing an atom in a tweezer handoff (x-axis). Each data point is an average over 400 initial configurations. The target is a $12\times12$ square grid with unit filling inside a $16\times16$ array with $60\%$ initial loading probability.
\textbf{c}, Ratio of Hungarian success probability $r_\text{H}$ to Balance and Compact success probability $r_{\text{BC}}$, indicating regions in which one algorithm is more effective. Contour lines have been smoothed for clarity using a Gaussian noise filter with $\sigma = 1$.
\textbf{d}, Linecut of the ratio at a vacuum-limited lifetime $t_\text{v}=13.90~\text{s}$, as indicated by the dash-dotted line in \textbf{c}. Error bars represent shot noise.}
\label{fig3}
\end{figure}

Subsequently, we extend our analysis to realistic experimental conditions. We compare two algorithms: the single-tweezer Hungarian algorithm (which minimizes transport distance) and the parallelized Balance and Compact algorithm (which aims to minimize rearrangement time), which we implement for tweezer arrays (Fig.~\ref{fig3}). 
We simultaneously sweep over a range of typical values for vacuum-limited lifetimes ($2 - 200~\text{s}$) and probabilities of losing an atom in a tweezer handoff ($0.05-0.5\%$), and compare the success probability of preparing a $12\times12$ fully filled configuration in a $16\times16$ array of tweezers with $60\%$ loading probability.

We find that the Hungarian is advantageous when the dominant sources of atom loss are tweezer errors. Since the Hungarian method minimizes the total transport distance by moving atoms one at a time, the number of pickup and putdown operations should be roughly $N$. In contrast, the multi-step structure of Balance and Compact may cause individual atoms to be picked up and moved many times. This extra overhead in the number of trap transfers makes the success rate decay faster with growing trap transfer error.

Conversely, we see that Balance and Compact is advantageous when vacuum-limited loss dominates (Fig.~\ref{fig3}c,d), as its high degree of parallelization enables sublinear time-scaling, whereas the absence of parallelization in the Hungarian means that its time-scaling goes super-linearly (see Table~\ref{scaling}). 
The ratio of success rates between the algorithms is shown for the entire sweep in Fig.~\ref{fig3}c, and a linecut for a fixed vacuum-limited lifetime of $13.90~\text{s}$ is shown in Fig.~\ref{fig3}d.

\section{Development and Benchmarking of heuristic dual-species algorithms}
\label{sec:dual}
Another relevant and lesser-explored problem is the development of algorithms that rearrange multiple atomic species inside the same array. 
Arrays comprised of rubidium and cesium \cite{Singh2022, Singh2023, Anand2024, Fang2025, White2026} and arrays with multiple isotopes of rubidium \cite{Sheng2022} have recently been experimentally realized, which motivates the search for an efficient dual-species rearrangement algorithm.

\begin{figure*}[t!]
\setlength{\tabcolsep}{10pt}
\centering
\includegraphics[width=\textwidth]
{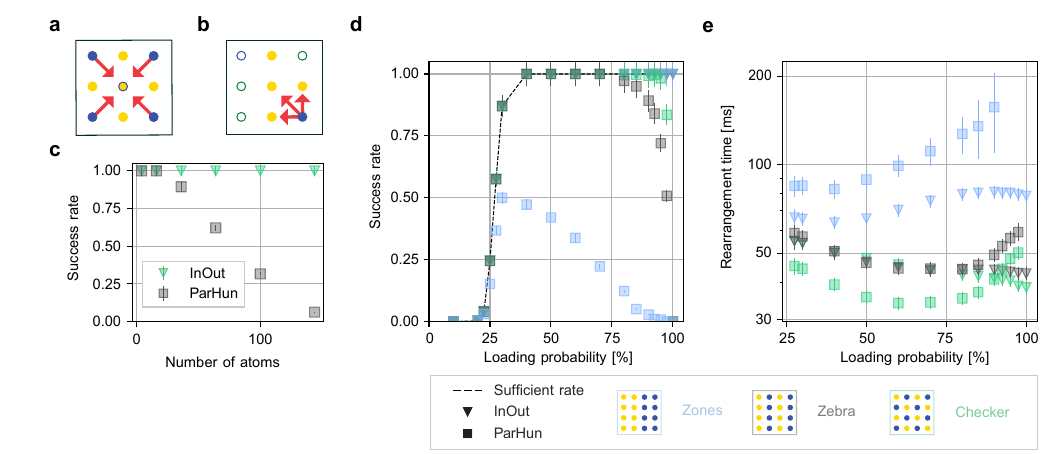}
\caption{
\textbf{Challenges of dual-species rearrangement and benchmarking of heuristic algorithms.}
\textbf{a}, \textbf{b}, Examples of `blocked' configurations, where blue atoms cannot move to target sites (marked by blue rings) due to the presence of (\textbf{a}) a yellow atom in the target site or (\textbf{b}) yellow atoms obstructing paths between the source atom and the target site.
\textbf{c}, Algorithm comparison performed on a 20 $\times$ 20 reservoir with a fixed loading probability of 60\%. While the InsideOut algorithm succeeds in preparing the target `Zones' configuration for all attempted initial configurations, a dual-species extension of ParHungarian encounters more blocked configurations as the target size increases.
\textbf{d}, \textbf{e}, Comparative benchmarking of InsideOut (triangular markers) and ParHungarian (square markers) for three target configurations: separate zones (blue), zebra stripes (gray), and checkerboard (green). All targets are $10\times10$ patterns inside a $20\times20$ array, and each data point is an average over 400 initial configurations.
\textbf{d}, Dependence of algorithm success rate on initial loading probability. The dashed black line represents the fraction of initial configurations with sufficient atom number to prepare the target configuration.
\textbf{e}, Rearrangement time under a realistic timing model (see  \ref{sec:timing_models} for details) plotted on a logarithmic vertical axis. ParHungarian markers are omitted for data points with 10 or fewer successful runs. Error bars represent shot noise.}
\label{fig5}
\end{figure*}

However, the task of dual-species rearrangement is fundamentally more complex than the single-species problem: whereas in single-species rearrangement the exchange of any two atoms does not change the state of the array nor the minimum number of moves to prepare some target configuration, in dual-species such an exchange may be nontrivial, and pairs of exchange operations do not necessarily commute.
While this observation may appear obvious, it follows that there exist detrimental configurations in which an atom is effectively `blocked' by neighboring atoms of the opposite species (Fig. \ref{fig5}a,b). 
Navigating such obstacles without ejecting atoms can introduce large time overheads.

Moreover, the existence of blocked configurations also raises the question of what the optimal experimental conditions are for dual-species rearrangement. 
In the single-species setting, an advantageous strategy is to increase the stochastic loading probability through experimental techniques including $\lambda$-enhanced gray molasses and dark-state enhanced loading \cite{BrownThiele2019,Shaw2023}.
Whereas here higher loading probability reduces the number of empty target sites and subsequently the number of moves required to fill the remaining target sites, in the dual-species setting the likelihood of developing blocked configurations also increases. This tradeoff naturally invites the question of whether there exists an optimal loading probability, and if so, whether it is consistent across all dual-species algorithms or varies with the choice of algorithm and/or target configuration.

While some initial algorithms have been proposed and realized experimentally to mitigate the challenges of dual-species rearrangement \cite{Wei2025,Sheng2022,Tao2022,Nakamura2024}, to the best of our knowledge, no algorithm has demonstrated the ability to prepare an arbitrary target configuration from any random initial configuration (with sufficient atom numbers and in the limit of large system size, i.e. an arbitrarily large reservoir) without moving atoms in the space between rows or columns. This constraint becomes important to minimize atom loss in dense arrays, as tweezers which move very close to each other may experience beating due to the small frequency shifts imparted by the AODs and risk being heated out of their traps. 

\textbf{Designing an efficient dual-species algorithm}.
We develop and benchmark a simple heuristic algorithm, InsideOut, which is able to prepare arbitrary target configurations with a near-unity success rate.
InsideOut is designed to circumvent such `blocked' configurations, and operates on a simple, layer-by-layer principle: first, it prepares the middle $2\times 2$ sites of the target configuration, pushing outward any misplaced atoms in the process.
Then, it prepares the next `layer' (in this case, the loop of twelve sites enclosing the middle $2\times 2$), and iteratively repeats until the target configuration is realized.
If at any point an atom of one species is blocking the path of an atom of opposite species, it is simply pushed away from the target configuration in a `clearing' operation.

Such a layer-by-layer approach reduces the chance to create `blocked' configurations and ensures that previous layers will not be disturbed by such `clearing' operations, thus preventing the emergence of `blocked' configurations (see \ref{sec:inside_out} for a detailed description).

\textbf{Benchmarking}.
To verify the efficacy of InsideOut, we check its scaling with increasing system size, and benchmark against a naive dual-species extension of the single-species ParHungarian algorithm (See \ref{sec:par_hung_dual}). 
We observe that InsideOut maintains a near-unity success rate with increasing system size, while the success rate of ParHungarian decreases to under $50\%$ in preparing a $10\times10$ target array (Fig.~\ref{fig5}c).

Next, we extend our benchmarking to study how initial loading probability impacts the success rate and the time costs of the rearrangement process (Fig.~\ref{fig5}d,e).

As expected, the success rate for InsideOut remains at unity across all loading probabilities while the success rate of ParHungarian decays at high loading probabilities (Fig.~\ref{fig5}d). Interestingly, for the `zebra stripe' and `checkerboard' configurations, ParHungarian appears to display clear minima  in rearrangement time at loading probabilities of around 60\%, whereas the corresponding InsideOut times appear to monotonically decrease with increasing loading probability.

However, behavior of both algorithms varies significantly with the choice of target configuration and density. In sparse arrays with fewer excess atoms, ParHungarian is able to prepare high-entropy `zebra stripe' and `checkerboard' patterns slightly faster than InsideOut. However, this quickly breaks down in dense arrays with many excess atoms, or when preparing the low-entropy `separate zones' configuration  (where InsideOut is advantageous in both rearrangement time and success rate). In these regimes, the paths for ParHungarian become increasingly full of obstacles, resulting in more circuitous paths and an overall slowdown, whereas the parallelized `clearing' operations of InsideOut remove such obstacles efficiently.

%Moreover, the rearrangement time of InsideOut decreases monotonically, while ParHungarian displays minima which vary significantly with the target configuration. While considering this observation, it is important to note that ParHungarian features extensive move parallelization, whereas InsideOut only parallelizes 'clearing' move segments, in which multiple obstacles in a layer are pushed away from the center.

\section{Discussion}
\label{sec:Discussion}
While there has been much activity in the experimental community to develop and implement new rearrangement algorithms, state-of-the-art processors are approaching sizes where naive algorithms become inefficient, motivating the development of algorithms with time-optimal scaling and adaptability to various noise regimes.

However, the scarcity of open-source implementations in the literature and the absence of error modeling makes such development and the comparison of new algorithms practically difficult.
By developing \textit{atommovr}, we hope to create a common platform for the community to discuss, compare existing algorithms and develop new algorithms with time-optimal scaling.

Finally, we wish to highlight alternate approaches to the crossed-AOD mechanism and areas for further development.

\textbf{Moving tweezers with SLMs or holographic AODs}. 
SLMs, while typically used to generate a grid of `static' tweezers, can also be used to move atoms between sites, thus eliminating the need for a separate set of mobile tweezers \cite{Kim2016, Knottnerus2025, Kim2019, Lin2024}. 
The key advantage of this method is that it avoids the constraint of generating moving tweezers on a rectilinear grid, and can execute arbitrary combinations of moves in parallel.
    
However, in practice, this method is hindered by relatively slow response times (frame generation rates) typically on the order of 1 ms (1 kHz) intrinsically limited by physical properties of the SLM \cite{Jullien2020}, the computational cost of calculating the input parameters for the desired hologram at each frame (typically by solving the Gerchberg-Saxton algorithm) and the discretization of the moving process. Recent work \cite{Treptow2021} has proposed using AODs to generate phase- and amplitude-controlled holograms, which would mitigate the frame rate issue.
    
\textbf{Continuous AOD tone modeling}. A current limitation of the modeling framework is that time costs for trap transfers, acceleration, and deceleration are applied uniformly across all moves in a single move round. Such an approximation allows for AOD operations to be efficiently discretely parametrized, enabling efficient simulation and collision detection (see \ref{sec:aod-model}), but at the cost of being conservative in estimating rearrangement time in certain scenarios. 
    
Modeling AOD operations as continuous functions would mitigate this issue. Moreover, such a construction could be built upon to enable detailed error modeling of heating from atom transport - an especially relevant problem for continuous reloading experiments \cite{Norcia2024,Pause24, Gyger2024} - via comparison of different ramping profiles.

\textbf{Non-heuristic machine learning methods}. Recently, advances in machine learning have enabled widespread advances in certain computational tasks. 
In particular, reinforcement learning has been notably used to build engines in the highly complex games of Go and chess that have defeated top human players~\cite{AlphaGo, AlphaGoZero, AlphaZeroGeneral}.
    
Considering that the underlying structure of Go is very similar to the rearrangement problem (any board position can be represented in the same manner as a dual-species atom array, with each board position having either a single black piece, a single white piece, or no piece), this could be a promising area for future study. 
In particular, an implementation of a single-player version of AlphaZero \cite{Moerland2022} to solve the rearrangement problem would be an exciting approach. 
Moreover, learning-driven algorithms might be particularly fruitful in practice compared to fixed heuristics, as reinforcement learning models tune their strategies via direct feedback from their environment.

\textbf{Acknowledgements.} We gratefully thank Shraddha Anand, Will Eckner, Andy Goldschmidt, Kevin Singh, Joanna Lis, and Mariesa Teo for stimulating discussions and Nayana Tiwari and Noah Glachman for insight in developing the code framework.

We acknowledge funding from the Office of Naval Research (Grant No. N00014-23-1-2540), the Air Force Office of Scientific Research (Grant No. FA9550-21-1-0209) and the Army Research Office
(Grant No. W911NF2410388). 
N.K.H. acknowledges support from the National Science Foundation Graduate Research Fellowship under Grant No. DGE-2040434. 
B.Y.C. acknowledges support from UChicago-Taiwan Student Exchange (UCTS) fellowship, National Taiwan University Fu Bell Scholarship, National Taiwan University College of Science Travel Grants and Chuan-Pu Lee Memorial Scholarship.

Correspondence and requests for materials should be directed to hannes.bernien@uibk.ac.at.

\textbf{Author Contributions.} N.K.H. and B.Y.C. built the code framework and developed the Parallel Hungarian algorithm and dual species algorithms. B.B. and H.B. offered crucial insight into the experimental constraints and simulation features. H.B. supervised the project. All authors read and revised the manuscript.

We acknowledge use of Large Language Models (LLMs) developed by Anthropic and OpenAI in writing and refactoring code.

%=========================
\let\oldaddcontentsline\addcontentsline
\renewcommand{\addcontentsline}[3]{}
\bibliographystyle{quantum}   % or plainnat
\bibliography{references}

\begin{thebibliography}{10}

\bibitem{Lee2016}
W.~Lee, H.~Kim, and J.~Ahn.
\newblock ``Three-dimensional rearrangement of single atoms using actively controlled optical microtraps''.
\newblock \href{https://dx.doi.org/10.1364/OE.24.009816}{Opt. Express {\bf 24}, 9816--9825}~(2016).

\bibitem{Barredo2016}
D.~Barredo, S.~de~Léséleuc, V.~Lienhard, T.~Lahaye, and A.~Browaeys.
\newblock ``An atom-by-atom assembler of defect-free arbitrary two-dimensional atomic arrays''.
\newblock \href{https://dx.doi.org/10.1126/science.aah3778}{Science {\bf 354}, 1021--1023}~(2016).

\bibitem{Endres2016}
M.~Endres, H.~Bernien, A.~Keesling, H.~Levine, E.~R. Anschuetz, A.~Krajenbrink, C.~Senko, V.~Vuleti{\'c}, M.~Greiner, and M.~D. Lukin.
\newblock ``Atom-by-atom assembly of defect-free one-dimensional cold atom arrays''.
\newblock \href{https://dx.doi.org/10.1126/science.aah3752}{Science {\bf 354}, 1024--1027}~(2016).

\bibitem{Kim2016}
H.~Kim, W.~Lee, H.~Lee, H.~Jo, Y.~Song, and J.~Ahn.
\newblock ``In situ single-atom array synthesis using dynamic holographic optical tweezers''.
\newblock \href{https://dx.doi.org/10.1038/ncomms13317}{Nature Communications {\bf 7}, 13317}~(2016).

\bibitem{Graham2022}
T.~M. Graham, Y.~Song, J.~Scott, C.~Poole, L.~Phuttitarn, K.~Jooya, P.~Eichler, X.~Jiang, A.~Marra, B.~Grinkemeyer, M.~Kwon, M.~Ebert, J.~Cherek, M.~T. Lichtman, M.~Gillette, J.~Gilbert, D.~Bowman, T.~Ballance, C.~Campbell, E.~D. Dahl, O.~Crawford, N.~S. Blunt, B.~Rogers, T.~Noel, and M.~Saffman.
\newblock ``Multi-qubit entanglement and algorithms on a neutral-atom quantum computer''.
\newblock \href{https://dx.doi.org/10.1038/s41586-022-04603-6}{Nature {\bf 604}, 457--462}~(2022).

\bibitem{Bluvstein2022}
D.~Bluvstein, H.~Levine, G.~Semeghini, T.~T. Wang, S.~Ebadi, M.~Kalinowski, A.~Keesling, N.~Maskara, H.~Pichler, M.~Greiner, V.~Vuleti{\'c}, and M.~D. Lukin.
\newblock ``A quantum processor based on coherent transport of entangled atom arrays''.
\newblock \href{https://dx.doi.org/10.1038/s41586-022-04592-6}{Nature {\bf 604}, 451--456}~(2022).

\bibitem{Reichardt2024}
B.~W. Reichardt et~al.
\newblock ``Logical computation demonstrated with a neutral atom quantum processor''.
\newblock \href{https://doi.org/10.48550/arXiv.2411.11822}{arXiv:2411.11822}~(2024).

\bibitem{Bluvstein2024}
D.~Bluvstein et~al.
\newblock ``Logical quantum processor based on reconfigurable atom arrays''.
\newblock \href{https://dx.doi.org/10.1038/s41586-023-06927-3}{Nature {\bf 626}, 58--65}~(2024).

\bibitem{Zhang2025}
B.~Zhang, G.~Liu, G.~Bornet, S.~P. Horvath, P.~Peng, S.~Ma, S.~Huang, S.~Puri, and J.~D. Thompson.
\newblock ``Logical qubits with erasure conversion using metastable neutral atoms''.
\newblock \href{https://dx.doi.org/10.1038/s41567-026-03309-0}{Nature Physics {\bf 22}, 910--916}~(2026).

\bibitem{Xu2024}
Q.~Xu, J.~P. Bonilla~Ataides, C.~A. Pattison, N.~Raveendran, D.~Bluvstein, J.~Wurtz, B.~Vasić, M.~D. Lukin, L.~Jiang, and H.~Zhou.
\newblock ``Constant-overhead fault-tolerant quantum computation with reconfigurable atom arrays''.
\newblock \href{https://dx.doi.org/10.1038/s41567-024-02479-z}{Nature Physics {\bf 20}, 1084--1090}~(2024).

\bibitem{Viszlai2024}
J.~Viszlai, W.~Yang, S.~F. Lin, J.~Liu, N.~Nottingham, J.~M. Baker, and F.~T. Chong.
\newblock ``Matching generalized-bicycle codes to neutral atoms for low-overhead fault-tolerance''.
\newblock In 2025 IEEE International Conference on Quantum Computing and Engineering (QCE).
\newblock \href{https://dx.doi.org/10.1109/QCE65121.2025.00080}{Volume~01, pages 688--699}.
\newblock ~(2025).

\bibitem{Poole2024}
C.~Poole, T.~M. Graham, M.~A. Perlin, M.~Otten, and M.~Saffman.
\newblock ``Architecture for fast implementation of quantum low-density parity-check codes with optimized {R}ydberg gates''.
\newblock \href{https://dx.doi.org/10.1103/PhysRevA.111.022433}{Phys. Rev. A {\bf 111}, 022433}~(2025).

\bibitem{B&Ctheory}
D.~S. Weiss, J.~Vala, A.~V. Thapliyal, S.~Myrgren, U.~Vazirani, and K.~B. Whaley.
\newblock ``Another way to approach zero entropy for a finite system of atoms''.
\newblock \href{https://dx.doi.org/10.1103/PhysRevA.70.040302}{Phys. Rev. A {\bf 70}, 040302}~(2004).

\bibitem{shaw2025erasure}
A.~L. Shaw, P.~Scholl, R.~Finkelstein, R.~B. Tsai, J.~Choi, and M.~Endres.
\newblock ``Erasure cooling, control, and hyperentanglement of motion in optical tweezers''.
\newblock \href{https://dx.doi.org/10.1126/science.adn2618}{Science {\bf 388}, 845--849}~(2025).

\bibitem{kumar2018sorting}
A.~Kumar, T.~Wu, F.~Giraldo, and D.~S. Weiss.
\newblock ``Sorting ultracold atoms in a three-dimensional optical lattice in a realization of {M}axwell’s demon''.
\newblock \href{https://dx.doi.org/10.1038/s41586-018-0458-7}{Nature {\bf 561}, 83--87}~(2018).

\bibitem{Schymik2020}
K.~Schymik, V.~Lienhard, D.~Barredo, P.~Scholl, H.~Williams, A.~Browaeys, and T.~Lahaye.
\newblock ``Enhanced atom-by-atom assembly of arbitrary tweezer arrays''.
\newblock \href{https://dx.doi.org/10.1103/PhysRevA.102.063107}{Phys. Rev. A {\bf 102}, 063107}~(2020).

\bibitem{Zhang2024}
Z.~Zhang, T.~Hsu, T.~Tan, D.~H. Slichter, A.~M. Kaufman, M.~Marinelli, and C.~A. Regal.
\newblock ``High optical access cryogenic system for {R}ydberg atom arrays with a 3000-second trap lifetime''.
\newblock \href{https://dx.doi.org/10.1103/PRXQuantum.6.020337}{PRX Quantum {\bf 6}, 020337}~(2025).

\bibitem{pichard2024rearrangement}
G.~Pichard, D.~Lim, \'E. Bloch, J.~Vaneecloo, L.~Bourachot, G.~Both, G.~M\'eriaux, S.~Dutartre, R.~Hostein, J.~Paris, B.~Ximenez, A.~Signoles, A.~Browaeys, T.~Lahaye, and D.~Dreon.
\newblock ``Rearrangement of individual atoms in a 2000-site optical-tweezer array at cryogenic temperatures''.
\newblock \href{https://dx.doi.org/10.1103/PhysRevApplied.22.024073}{Phys. Rev. Appl. {\bf 22}, 024073}~(2024).

\bibitem{Manetsch2024}
H.~J. Manetsch, G.~Nomura, E.~Bataille, X.~Lv, K.~H. Leung, and M.~Endres.
\newblock ``A tweezer array with 6,100 highly coherent atomic qubits''.
\newblock \href{https://dx.doi.org/10.1038/s41586-025-09641-4}{Nature {\bf 647}, 60--67}~(2025).

\bibitem{Norcia2024}
M.~A. Norcia et~al.
\newblock ``Iterative assembly of ${}^{171}$$\mathrm{Yb}$ atom arrays with cavity-enhanced optical lattices''.
\newblock \href{https://dx.doi.org/10.1103/PRXQuantum.5.030316}{PRX Quantum {\bf 5}, 030316}~(2024).

\bibitem{Pause24}
L.~Pause, L.~Sturm, M.~Mittenb\"{u}hler, S.~Amann, T.~Preuschoff, D.~Sch\"{a}ffner, M.~Schlosser, and G.~Birkl.
\newblock ``Supercharged two-dimensional tweezer array with more than 1000 atomic qubits''.
\newblock \href{https://dx.doi.org/10.1364/OPTICA.513551}{Optica {\bf 11}, 222--226}~(2024).

\bibitem{Gyger2024}
F.~Gyger, M.~Ammenwerth, R.~Tao, H.~Timme, S.~Snigirev, I.~Bloch, and J.~Zeiher.
\newblock ``Continuous operation of large-scale atom arrays in optical lattices''.
\newblock \href{https://dx.doi.org/10.1103/PhysRevResearch.6.033104}{Phys. Rev. Res. {\bf 6}, 033104}~(2024).

\bibitem{Treptow2021}
D.~Treptow, R.~Bola, E.~Martín-Badosa, and M.~Montes-Usategui.
\newblock ``Artifact-free holographic light shaping through moving acousto-optic holograms''.
\newblock \href{https://dx.doi.org/10.1038/s41598-021-00332-4}{Scientific Reports {\bf 11}, 21261}~(2021).

\bibitem{Knottnerus2025}
I.~H.~A. Knottnerus, Y.~C. Tseng, A.~Urech, R.~J.~C. Spreeuw, and F.~Schreck.
\newblock ``{Parallel assembly of neutral atom arrays with an SLM using linear phase interpolation}''.
\newblock \href{https://dx.doi.org/10.21468/SciPostPhys.19.4.118}{SciPost Phys. {\bf 19}, 118}~(2025).

\bibitem{Wei2025}
M.~Wei, K.~Wang, J.~Hou, Y.~Chen, P.~Xu, J.~Zhuang, R.~Guo, M.~Liu, J.~Wang, X.~He, and M.~Zhan.
\newblock ``Enhanced atom-by-atom assembly of defect-free two-dimensional mixed-species atomic arrays''.
\newblock \href{https://dx.doi.org/10.1103/4hhn-pjbj}{Phys. Rev. Appl. {\bf 25}, 034009}~(2026).

\bibitem{Konig1916}
D.~König.
\newblock ``Über {G}raphen und ihre {A}nwendung auf {D}eterminantentheorie und {M}engenlehre''.
\newblock \href{https://dx.doi.org/10.1007/BF01456961}{Mathematische Annalen {\bf 77}, 453--465}~(1916).

\bibitem{Egervary1931}
J.~Egerv\'{a}ry.
\newblock ``Matrixok kombinatorius tulajdons\'{a}gair\'{o}l''.
\newblock \href{https://real-j.mtak.hu/7307/}{Mat.\ Fiz.\ Lapok \textbf{38}, 16--28}~(1931).

\bibitem{Kuhn1955}
H.~W. Kuhn.
\newblock ``The {H}ungarian method for the assignment problem''.
\newblock \href{https://dx.doi.org/10.1002/nav.3800020109}{Naval Research Logistics Quarterly {\bf 2}, 83--97}~(1955).

\bibitem{Lee2017}
W.~Lee, H.~Kim, and J.~Ahn.
\newblock ``Defect-free atomic array formation using the {H}ungarian matching algorithm''.
\newblock \href{https://dx.doi.org/10.1103/PhysRevA.95.053424}{Phys. Rev. A {\bf 95}, 053424}~(2017).

\bibitem{Tian2023}
W.~Tian, W.~J. Wee, A.~Qu, B.~J.~M. Lim, P.~R. Datla, V.~P.~W. Koh, and H.~Loh.
\newblock ``Parallel assembly of arbitrary defect-free atom arrays with a multitweezer algorithm''.
\newblock \href{https://dx.doi.org/10.1103/PhysRevApplied.19.034048}{Phys. Rev. Appl. {\bf 19}, 034048}~(2023).

\bibitem{zhang2025fast}
Y.~Zhang, Z.~Zhang, G.~Zhang, Z.~Zhang, Y.~Chen, Y.~Li, W.~Liu, J.~Wu, V.~Sovkov, and J.~Ma.
\newblock ``A fast rearrangement method for defect-free atom arrays''.
\newblock \href{https://dx.doi.org/10.3390/photonics12020117}{Photonics {\bf 12}, 117}~(2025).

\bibitem{Cimring2023}
B.~Cimring, R.~El~Sabeh, M.~Bacvanski, S.~Maaz, I.~El~Hajj, N.~Nishimura, A.~E. Mouawad, and A.~Cooper.
\newblock ``Efficient algorithms to solve atom reconfiguration problems. {I}. redistribution-reconfiguration algorithm''.
\newblock \href{https://dx.doi.org/10.1103/PhysRevA.108.023107}{Phys. Rev. A {\bf 108}, 023107}~(2023).

\bibitem{B&C}
J.~Vala, A.~V. Thapliyal, S.~Myrgren, U.~Vazirani, D.~S. Weiss, and K.~B. Whaley.
\newblock ``Perfect pattern formation of neutral atoms in an addressable optical lattice''.
\newblock \href{https://dx.doi.org/10.1103/PhysRevA.71.032324}{Phys. Rev. A {\bf 71}, 032324}~(2005).

\bibitem{Wang2023}
S.~Wang, W.~Zhang, T.~Zhang, S.~Mei, Y.~Wang, J.~Hu, and W.~Chen.
\newblock ``Accelerating the assembly of defect-free atomic arrays with maximum parallelisms''.
\newblock \href{https://dx.doi.org/10.1103/PhysRevApplied.19.054032}{Phys. Rev. Appl. {\bf 19}, 054032}~(2023).

\bibitem{PPSU}
I.~Panagiotas, G.~Pichon, S.~Singh, and B.~U\c{c}ar.
\newblock ``{Engineering Fast Algorithms for the Bottleneck Matching Problem}''.
\newblock In 31st Annual European Symposium on Algorithms (ESA 2023).
\newblock \href{https://dx.doi.org/10.4230/LIPIcs.ESA.2023.87}{Volume 274 of Leibniz International Proceedings in Informatics (LIPIcs), pages 87:1--87:15}.
\newblock Dagstuhl, Germany~(2023). Schloss Dagstuhl -- Leibniz-Zentrum f{\"u}r Informatik.

\bibitem{Burkard1980}
R.~E. Burkard and U.~Derigs.
\newblock ``The bottleneck matching problem''.
\newblock In Assignment and Matching Problems: Solution Methods with FORTRAN-Programs.
\newblock \href{https://dx.doi.org/10.1007/978-3-642-51576-7_5}{Pages 60--71}.
\newblock Springer, Berlin, Heidelberg~(1980).

\bibitem{Singh2022}
K.~Singh, S.~Anand, A.~Pocklington, J.~T. Kemp, and H.~Bernien.
\newblock ``Dual-element, two-dimensional atom array with continuous-mode operation''.
\newblock \href{https://dx.doi.org/10.1103/PhysRevX.12.011040}{Phys. Rev. X {\bf 12}, 011040}~(2022).

\bibitem{Singh2023}
K.~Singh, C.~E. Bradley, S.~Anand, V.~Ramesh, R.~White, and H.~Bernien.
\newblock ``Mid-circuit correction of correlated phase errors using an array of spectator qubits''.
\newblock \href{https://dx.doi.org/10.1126/science.ade5337}{Science {\bf 380}, 1265--1269}~(2023).

\bibitem{Anand2024}
S.~Anand, C.~E. Bradley, R.~White, V.~Ramesh, K.~Singh, and H.~Bernien.
\newblock ``A dual-species {R}ydberg array''.
\newblock \href{https://dx.doi.org/10.1038/s41567-024-02638-2}{Nature Physics {\bf 20}, 1744--1750}~(2024).

\bibitem{Fang2025}
C.~Fang, J.~Miles, J.~Goldwin, M.~Lichtman, M.~Gillette, M.~Bergdolt, S.~Deshpande, S.~A. Norrell, P.~Huft, M.~A. Kats, and M.~Saffman.
\newblock ``Interleaved dual-species arrays of single atoms using a passive optical element and one trapping laser''.
\newblock \href{https://dx.doi.org/10.1126/sciadv.adw4166}{Science Advances {\bf 11}, eadw4166}~(2025).

\bibitem{White2026}
R.~White, V.~Ramesh, A.~Impertro, S.~Anand, F.~Cesa, G.~Giudici, T.~Iadecola, H.~Pichler, and H.~Bernien.
\newblock ``Quantum cellular automata on a dual-species {R}ydberg processor''.
\newblock \href{https://doi.org/10.48550/arXiv.2601.16257}{arXiv:2601.16257}~(2026).

\bibitem{Sheng2022}
C.~Sheng, J.~Hou, X.~He, K.~Wang, R.~Guo, J.~Zhuang, B.~Mamat, P.~Xu, M.~Liu, J.~Wang, and M.~Zhan.
\newblock ``Defect-free arbitrary-geometry assembly of mixed-species atom arrays''.
\newblock \href{https://dx.doi.org/10.1103/PhysRevLett.128.083202}{Phys. Rev. Lett. {\bf 128}, 083202}~(2022).

\bibitem{BrownThiele2019}
M.~O. Brown, T.~Thiele, C.~Kiehl, T.-W. Hsu, and C.~A. Regal.
\newblock ``Gray-molasses optical-tweezer loading: Controlling collisions for scaling atom-array assembly''.
\newblock \href{https://dx.doi.org/10.1103/PhysRevX.9.011057}{Phys. Rev. X {\bf 9}, 011057}~(2019).

\bibitem{Shaw2023}
A.~L. Shaw, P.~Scholl, R.~Finkelstein, I.~S. Madjarov, B.~Grinkemeyer, and M.~Endres.
\newblock ``Dark-state enhanced loading of an optical tweezer array''.
\newblock \href{https://dx.doi.org/10.1103/PhysRevLett.130.193402}{Phys. Rev. Lett. {\bf 130}, 193402}~(2023).

\bibitem{Tao2022}
Z.~Tao, L.~Yu, P.~Xu, J.~Hou, X.~He, and M.~Zhan.
\newblock ``Efficient two-dimensional defect-free dual-species atom arrays rearrangement algorithm with near-fewest atom moves''.
\newblock \href{https://dx.doi.org/10.1088/0256-307X/39/8/083701}{Chinese Physics Letters {\bf 39}, 083701}~(2022).

\bibitem{Nakamura2024}
Y.~Nakamura, T.~Kusano, R.~Yokoyama, K.~Saito, K.~Higashi, N.~Ozawa, T.~Takano, Y.~Takasu, and Y.~Takahashi.
\newblock ``Hybrid atom tweezer array of nuclear spin and optical clock qubits''.
\newblock \href{https://dx.doi.org/10.1103/PhysRevX.14.041062}{Phys. Rev. X {\bf 14}, 041062}~(2024).

\bibitem{Kim2019}
H.~Kim, M.~Kim, W.~Lee, and J.~Ahn.
\newblock ``Gerchberg-{S}axton algorithm for fast and efficient atom rearrangement in optical tweezer traps''.
\newblock \href{https://dx.doi.org/10.1364/OE.27.002184}{Opt. Express {\bf 27}, 2184--2196}~(2019).

\bibitem{Lin2024}
R.~Lin et~al.
\newblock ``A{I}-enabled parallel assembly of thousands of defect-free neutral atom arrays''.
\newblock \href{https://dx.doi.org/10.1103/2ym8-vs82}{Phys. Rev. Lett. {\bf 135}, 060602}~(2025).

\bibitem{Jullien2020}
A.~Jullien.
\newblock ``Spatial light modulators''.
\newblock \href{https://dx.doi.org/10.1051/photon/202010159}{Photoniques {\bf 101}, 59--64}~(2020).

\bibitem{AlphaGo}
D.~Silver et~al.
\newblock ``Mastering the game of {G}o with deep neural networks and tree search''.
\newblock \href{https://dx.doi.org/10.1038/nature16961}{Nature {\bf 529}, 484--489}~(2016).

\bibitem{AlphaGoZero}
D.~Silver et~al.
\newblock ``Mastering the game of {G}o without human knowledge''.
\newblock \href{https://dx.doi.org/10.1038/nature24270}{Nature {\bf 550}, 354--359}~(2017).

\bibitem{AlphaZeroGeneral}
D.~Silver et~al.
\newblock ``A general reinforcement learning algorithm that masters chess, shogi, and {G}o through self-play''.
\newblock \href{https://dx.doi.org/10.1126/science.aar6404}{Science {\bf 362}, 1140--1144}~(2018).

\bibitem{Moerland2022}
T.~Moerland.
\newblock ``Single player {A}lpha {Z}ero implementation''.
\newblock \url{https://github.com/tmoer/alphazero_singleplayer}~(2022).
\newblock Accessed: 2025-07-11.

\end{thebibliography}
\let\addcontentsline\oldaddcontentsline

\clearpage
\pagebreak

\setcounter{page}{1}
\setcounter{equation}{0}
\setcounter{figure}{0}

\appendix

\renewcommand{\thesection}{Appendix \Alph{section}}
\renewcommand{\theequation}{S.\arabic{equation}}
\renewcommand{\thefigure}{S\arabic{figure}}
\renewcommand{\thepage}{S\arabic{page}}

\onecolumngrid

\begin{center}
    {\large\textbf{
        Supplementary Materials for\\
        ``atommovr: An open-source simulation framework for rearrangement in atomic arrays''
    }}

    \vspace{0.5cm}

    Nikhil K. Harle\textsuperscript{1,2,*},
    Bo-Yu Chen\textsuperscript{1,3,4,*},
    Bob Bao\textsuperscript{5,6},
    and Hannes Bernien\textsuperscript{1,3,7}

    \vspace{0.25cm}

    \textit{
        \textsuperscript{1}\pme\\
        \textsuperscript{2}\jila\\
        \textsuperscript{3}\iqoqi\\
        \textsuperscript{4}\ntu\\
        \textsuperscript{5}\psd\\
        \textsuperscript{6}\harvardqse\\
        \textsuperscript{7}\uibk
    }

    \vspace{0.2cm}

    \textsuperscript{*}These authors contributed equally to this work.
\end{center}

\vspace{0.5cm}

\section{Physical assumptions and timing models}
\subsection{Parametrization and legality of AOD move segments}
\label{sec:aod-model}
We parametrize each AOD move segment with two vectors, $c_\text{row}$ and $c_\text{col}$, of length $n_\text{rows}$ and $n_\text{cols}$, respectively, where each entry describes the status of the RF tone corresponding to a particular row or column. We adopt the following discrete encoding: 

$c^i_\text{row (col)} = 0$ if the RF tone corresponding to the $i$-th row (column) is inactive; 

$c^i_\text{row (col)} = 1$ if the RF tone corresponding to the $i$-th row (column) is active and static; 

$c^i_\text{row (col)} = 2$ if the RF tone corresponding to the $i$-th row (column) is active and ramped to the frequency of the $\left(i+1\right)$th row (column);

$c^i_\text{row (col)} = 3$ if the RF tone corresponding to the $i$-th row (column) is active and ramped to the frequency of the $\left(i-1\right)$th row (column). 

An AOD tweezer is present at the $i$-th row and $j$-th column if $c^i_\text{row}, c^j_\text{col} \neq 0$.

It is important to note that this parametrization includes AOD move segments where tweezers collide with one another (i.e. illegal segments). For example, take the AOD move segment described by $c_\text{row} = \left[2, 0, 3\right]$ and $c_\text{col} = \left[0, 1, 0\right]$. This describes two tweezers at the intersections of the first and third rows with the second column, that both move to the intersection of the second row with the second column. Since the tweezers converge to the same spot, any atoms they carry are assumed to be heated out of the traps either due to light-assisted collisions (if both tweezers are occupied) or beating between the converging tones (if only one tweezer is occupied).

Concretely, we define an AOD tweezer to be legal iff (1) it does not intersect with another AOD tweezer at any point in its trajectory and (2) if carrying an atom, it does not intersect with an occupied SLM tweezer. In the case of (1), both atoms in intersecting tweezers are assumed to be deterministically lost. In the case of (2), both the atom in the AOD tweezer and the atom in the SLM tweezer are assumed to be deterministically lost.

\subsection{Timing models}
\label{sec:timing_models}
The error models we use in \textit{atommovr} track time costs associated with trap transfers (i.e. transferring an atom from a static SLM trap to a mobile AOD trap (``pickup") and vice versa (``putdown")) and move distance, along with additional time costs for gently accelerating and decelerating atoms from/to rest.

In this work, we use two timing models: a naive model which neglects time costs for trap transfers and acceleration/deceleration, and a detailed model which takes into account typical atom transfer trajectories. 
Here we define $s$ as the array spacing, $v_\text{avg}$ as the average move velocity, and $t_\text{avg}=\nicefrac{s}{v_\text{avg}}$ as the average single-site move time.

In the naive model, the time for an AOD move segment is equal to the maximum distance $d_{\max}$ traveled by an atom:
\begin{equation}
    t_\text{naive} = \frac{d_{\max}}{v_\text{avg}}
\end{equation}
where $d_{\max}=s$ if the AOD move segment contains only horizontal or vertical motion and $d_{\max}=\sqrt{2}s$ if the move segment involves diagonal displacement.

In the detailed model, we parametrize the move-distance-dependent time cost and the acceleration/deceleration time costs through considering the adiabatic sine model for atom transport used in \cite{Manetsch2024}:
\begin{equation}
\xi(u)=\frac{1}{\pi}\sin(\pi u)+u,\qquad u,\xi\in[-1,1],
\end{equation}
where $\xi$ and $u$ are normalized position and time variables mapped from displacement and time variables $x$ and $t$.
By rescaling $\xi$ and $u$, we obtain
%Taking $x(0) = 0$ and $x(t_\text{avg}=\nicefrac{s}{v_\text{avg}}) = s$ (we omit simple algebraic steps):
\begin{equation}
     x(t) = \frac{s}{2\pi}\sin\left(\frac{2\pi}{t_\text{avg}}\left(t-\frac{t_\text{avg}}{2}\right)\right) + s\frac{t}{t_\text{avg}} 
\end{equation}
where now $t \in [0, t_\text{avg}]$ and $x \in [0, s]$.

For multi-site and diagonal moves of total distance $d$, we adopt a piecewise construction. The acceleration profile follows the single-site adiabatic sine profile for $x \in [0, \nicefrac{s}{2}]$, the atom remains at the maximum velocity $v_\text{max}$ to travel the middle interval $x\in[\nicefrac{s}{2}, d-\nicefrac{s}{2}]$, and the deceleration profile follows the latter half of the single-site adiabatic sine profile for $x\in[d-\nicefrac{s}{2}, d]$.

With this piecewise construction, we can calculate the extra time costs associated with acceleration and deceleration by returning to the single-site move case:
%\begin{equation}
%\begin{split}
%    t_\text{avg} & = \frac{s}{v_\text{avg}} \\
 %   & = \frac{s}{v_\text{max}} + t_\text{accel} + t_\text{decel} \\
 %   & = \frac{s}{v_\text{max}} + 2t_\text{accel}
%\end{split}
%\end{equation}
%where we take $t_\text{accel} = t_\text{decel}$.
\begin{equation}
\begin{split}
    t_\text{avg} & = \frac{s}{v_\text{avg}} \\
    & = \frac{s}{v_\text{max}} + t_\text{accel} + t_\text{decel},
\end{split}
\end{equation}
which implies
\begin{equation}
    t_\text{accel} + t_\text{decel} =s \left(\frac{1}{v_\text{avg}} - \frac{1}{v_\text{max}}\right)
\end{equation}
and subsequently, we can write the detailed time cost for a single AOD move segment to be
\begin{equation}
\label{s7}
\begin{split}
    t_\text{detailed} 
    &=\frac{d_\text{max}}{2v_\text{avg}} + t_\text{accel} \delta_\text{accel}+t_\text{decel}\delta_\text{decel}\\
    &+t_\text{transfer}(\delta_\text{putdown} +\delta_\text{pickup})
\end{split}
\end{equation}
where $\delta_{(*)}$ = 1 if a ($*$) event occurs for any tweezer in the AOD move segment and 0 otherwise, and $t_\text{transfer}$ is the time for a trap transfer between SLM and AOD tweezers.

In our simulations, taking $t_\text{accel} = t_\text{decel}$ and the fact that $v_\text{max} = 2v_\text{avg}$ for the adiabatic sine curve, Eq.~\ref{s7} is simplified as:

\begin{equation}
\begin{split}
    t_\text{detailed} 
    &=\frac{d_\text{max}}{2v_\text{avg}} + \frac{s}{4v_\text{avg}}( \delta_\text{accel}+\delta_\text{decel})\\
    &+t_\text{transfer}(\delta_\text{putdown} +\delta_\text{pickup})
\end{split}
\end{equation}

For the data presented in the paper, we take $s = 5~\mu\text{m}$ and $v_\text{avg} = 0.1~\nicefrac{\text{m}}{\text{s}}$, which in the detailed model sets the acceleration and deceleration times to $t_\text{accel} = t_\text{decel} =  12.5~\mu\text{s}$.
Following \cite{Manetsch2024, White2026}, we take $t_\text{transfer} = 200~\mu\text{s}$. 

\section{Proposed algorithms}

% \subsection{Generalized Balance (single-species)}
% \label{sec:gen_balance}
% To be filled.

\subsection{Parallel Hungarian (single-species)}
\label{sec:par_hung_single}
To explore the extent to which parallelization of moves can naively be tacked on to existing algorithms, we propose and build a variation of the \textit{distance}-optimal Hungarian algorithm, hereafter referred to as ParHungarian.
Visualization of the algorithm is shown in Fig.~\ref{fig:par_Hung_flow}.

\usetikzlibrary{arrows.meta,positioning,calc,shapes.geometric}

\begin{figure}[p]
\centering

\begin{tikzpicture}[
    scale=0.72,
    transform shape,
    font=\small,
    node distance=7mm and 16mm,
    flowarrow/.style={
        -Latex,
        line width=0.75pt
    },
    startstop/.style={
        rectangle,
        rounded corners=2mm,
        draw,
        minimum height=10mm,
        inner xsep=4mm,
        align=center
    },
    io/.style={
        trapezium,
        trapezium left angle=70,
        trapezium right angle=110,
        draw,
        minimum height=11mm,
        text width=78mm,
        align=center,
        inner ysep=2mm
    },
    process/.style={
        rectangle,
        draw,
        minimum height=11mm,
        text width=74mm,
        align=center,
        inner sep=2.5mm
    },
    decision/.style={
        diamond,
        draw,
        aspect=3.5,
        text width=62mm,
        align=center,
        inner sep=1.5mm
    },
    branchlabel/.style={
        font=\small,
        fill=white,
        inner sep=1pt
    }
]

% ------------------------------------------------
% Main vertical sequence
% ------------------------------------------------

\node (start) [startstop]
{Start rearrangement};

\node (load) [io, below=of start]
{
    Initial loading array
    $\mathbf{X}_0\in\{0,1\}^{m\times n}$
    \\
    Target pattern
    $\mathbf{g}\in\{0,1\}^{m\times n}$
};

\node (enough) [decision, below=10mm of load]
{
    Are there sufficient atoms to prepare
    the target configuration?
};

\node (sites) [process, below=11mm of enough]
{
    Find excess-atom sites and target vacancies,
    and generate their cost matrix $\mathbf{C}$
};

\node (assign) [process, below=of sites]
{
    Generate $N'$ atom--target pairs
    with minimum total distance
};

\node (paths) [process, below=of assign]
{
    Find the shortest path or subpaths for the $i$th pair,
    denoted by $\{p_\mu\}_{\mu=1}^{N}$
    \\
    Let
    $p_\mu:=\{m_{\mu t}\}_{t=1}^{L_\mu}$
    and
    $L:=\max_\mu \operatorname{Len}(p_\mu)$
};

\node (sorted) [decision, below=11mm of paths]
{
    Have all moves in every path been sorted?
};

\node (collect) [process, below=11mm of sorted]
{
    Remove the first element from each nonempty path
    and add the elements to a new list $q_\mu$.
    \\[-0.5mm]
    {\scriptsize
    \textsuperscript{a}A path may be empty because
    path lengths can differ.}
};

\node (parallel) [process, below=of collect]
{
    Check for collisions among moves in $q_\mu$.
    If a collision occurs, retain one move in $q_\mu$
    and return the others to their corresponding paths $p_\mu$.
    \\
    Parallelize the retained moves
    (Fig.~\ref{fig:make_paralell}).
};

\node (execute) [
    process,
    text width=90mm,
    below=of parallel
]
{
    Construct a sequence of operations
    $\widetilde{\epsilon}
    =\{\epsilon_\nu\}_{\nu=1}^{N'}$,
    with $N'\leq N$.
    \\
    Execute the moves and update the array:
    $\mathbf{X}_{t+N'}
    =\widetilde{\epsilon}(\mathbf{X}_t)$.
};

\node (return) [
    io,
    text width=58mm,
    below=of execute
]
{
    Return final array
    $\mathbf{X}_{\mathrm{final}}$
};

\node (end) [startstop, below=of return]
{
    Finish rearrangement
};

% ------------------------------------------------
% Main downward arrows
% ------------------------------------------------

\draw[flowarrow] (start) -- (load);

\draw[flowarrow] (load) -- (enough);

\draw[flowarrow]
    (enough)
    --
    node[branchlabel, left] {True}
    (sites);

\draw[flowarrow] (sites) -- (assign);

\draw[flowarrow] (assign) -- (paths);

\draw[flowarrow] (paths) -- (sorted);

\draw[flowarrow]
    (sorted)
    --
    node[branchlabel, left] {False}
    (collect);

\draw[flowarrow] (collect) -- (parallel);

\draw[flowarrow] (parallel) -- (execute);

%\draw[flowarrow] (execute) -- (return);

\draw[flowarrow] (return) -- (end);

% ------------------------------------------------
% False branch from atom-number decision
% ------------------------------------------------

\coordinate (outerRight)
    at ($(enough.east)+(32mm,0)$);

\draw[flowarrow]
    (enough.east)
    --
    node[branchlabel, above] {False}
    (outerRight)
    |-
    (return.east);

% ------------------------------------------------
% True branch from sorting decision
% ------------------------------------------------

\coordinate (innerRight)
    at ($(sorted.east)+(20mm,0)$);

\draw[flowarrow]
    (sorted.east)
    --
    node[branchlabel, above] {True}
    (innerRight)
    |-
    (return.east);

% ------------------------------------------------
% Loop from executed batch back to sorting decision
% ------------------------------------------------

\coordinate (loopLeft)
    at ($(execute.west)+(-20mm,0)$);

\draw[flowarrow]
    (execute.west)
    --
    (loopLeft)
    |-
    (sorted.west);

\end{tikzpicture}

\caption{Flowchart of the ParHungarian algorithm (single species).}
\label{fig:par_Hung_flow}
\end{figure}

\paragraph{Detailed description of ParHungarian.}
First, the algorithm checks if the number of atoms given by the initial random loading is sufficient to prepare the target pattern, i.e. $\sum_{ij} \mathbf{X}^{\text{init}}_{ij} \geq \sum_{ij}\mathbf{g}_{ij}$ (to keep notations clean, here we use $\mathbf{X}^\text{init}$ to represent $\mathbf{X}_0$);
if not, the algorithm stops immediately.
Similarly to the original Hungarian algorithm, ParHungarian generates a cost matrix $\mathbf{C}$ where the rows are target sites to be filled, i.e. site $(i,j)$ is a target site if $\mathbf{X}^\text{init}_{ij} = 0$ and $\mathbf{g}_{ij} = 1$, the columns are excess atoms, i.e. site $(i',j')$ contains an excess atom if $\mathbf{X}^\text{init}_{i'j'} = 1$ and $\mathbf{g}_{i'j'} = 0$, and the component $\mathbf{C}_{kl}$ is given by the Euclidean distance between the $k$th target site and the $l$th excess atom \cite{Lee2017}.
By solving the linear sum assignment problem corresponding to the cost matrix $\mathbf{C}$, ParHungarian assigns each target vacancy to an excess atom, denoted as a mapping $h: a \to t$, where $a$ and $t$ represent the set of excess atoms and the set of target vacancies, respectively.
According to the mapping relationship, ParHungarian generates $N'$ atom-target pairs, i.e. $\{a_\nu, t_\nu\}_{\nu=1}^{N'}$, where $N'$ is the size of set $a$.
Note that the number of excess atoms is always greater or equal to the number of target vacancies due to the initial sufficient atom check.

According to the assignments, ParHungarian finds the shortest direct path between each pair.
If no direct path exists between $a_\nu$ and $t_\nu$ due to the presence of obstacle atoms, the path is decomposed into a sequence of `domino' subpaths.
In this scheme, the final obstacle is first relocated to the target position $t_\nu$, after which each preceding obstacle successively occupies the position vacated by the following obstacle, until $a_\nu$ can be moved to the location of the first obstacle along the path.
% Therefore, the algorithm creates multiple segmant(s) of moves and moves atoms like triggering domino.
By the above process, ParHungarian generates $N$ paths, where $N$ is the sum of the number of direct paths and number of additional segments in the domino paths, denoted as $\{p_\mu\}_{\mu=1}^N$, where each path $p_\mu := \{m_{\mu t}\}_{t=1}^{L_\mu}$, $L_\mu := \text{Len}(p_\mu)$ is the length of the $\mu$th path, and $m_{\mu t}$ is the $t$th move element of the path $\mu$.
% Note that since these $N$ paths include direct paths and domino segmant paths, $N\geq N'$ always holds.

Here, we define $L$ as the length of the longest path, i.e. $L\coloneqq \max_\mu L_\mu$.
Firstly, ParHungarian removes the 1st moves in every path and adds them to a new list $q_1 :=\{m_{\mu 1}\}_{\mu=1}^N$.
Then, ParHungarian checks if there are two or more moves in $q_1$ whose simultaneous execution would result in a collision.
If such a case or cases are found, ParHungarian only keeps one move in this round of execution and puts the others back in their respective path lists.
The retained moves are subsequently sorted into a set of sequential `AOD moves' (Fig.~\ref{fig:make_paralell}).
% After checking no contradiction and finishing parallelization, ParHungarian moves the atoms and updates the atom array information.
We repeat this process, constructing subsequent lists $q_2, q_3, \ldots$, removing the first elements of the modified path lists, until all path lists $p_\mu$ are empty. After all moves have been sorted into sets of sequential `AOD moves', the rearrangement process completes.
% This process is repeated to construct the $2$nd, $3$rd, $\ldots$, $L$th lists $\{q_{2}, q_3, \ldots, q_L\}$ of moves, after which the rearrangement completes.
% Note that the length of each path might not be the same, so $m_\mu$ in some path could be empty and skipped by the algorithm.

\begin{figure}[p]
\centering

\begin{tikzpicture}[
    scale=0.72,
    transform shape,
    font=\small,
    node distance=7mm and 16mm,
    flowarrow/.style={
        -Latex,
        line width=0.75pt
    },
    startstop/.style={
        rectangle,
        rounded corners=2mm,
        draw,
        minimum height=10mm,
        inner xsep=4mm,
        align=center
    },
    io/.style={
        trapezium,
        trapezium left angle=70,
        trapezium right angle=110,
        draw,
        minimum height=11mm,
        text width=78mm,
        align=center,
        inner ysep=2mm
    },
    process/.style={
        rectangle,
        draw,
        minimum height=11mm,
        text width=74mm,
        align=center,
        inner sep=2.5mm
    },
    decision/.style={
        diamond,
        draw,
        aspect=3.5,
        text width=62mm,
        align=center,
        inner sep=1.5mm
    },
    branchlabel/.style={
        font=\small,
        fill=white,
        inner sep=1pt
    }
]

% ------------------------------------------------
% Main vertical sequence
% ------------------------------------------------

\node (start) [startstop]
{Start parallelization};

\node (input) [io, below=of start]
{
    A list of moves $\mathcal{M}=\{m_1,\ldots,m_M\}$
    \\
    Atom array $\mathbf{X}$
};

\node (nonempty) [decision, below=10mm of input]
{
    $\mathcal{M}$ is not empty
};

\node (init) [process, below=11mm of nonempty]
{
    Initialize temporary move queue
    $\widetilde{\epsilon}=\{m_i\}$;
    \\
    remove $m_i$ from $\mathcal{M}$;
    \\
    $i$ is the smallest index in $\mathcal{M}$
};

\node (iterate) [decision, below=11mm of init, text width=74mm]
{
    Iterate $j=2,\ldots,M'$ and $m_j\in\mathcal{M}$
};

\node (parstate) [process, below=11mm of iterate]
{
    Update array
    $\mathbf{X}'_{\mathrm{Par}}
    =\widetilde{\epsilon}(\mathbf{X})$
};

\node (seqstate) [process, below=of parstate, text width=84mm]
{
    Decompose $\widetilde{\epsilon}$ into a set of AOD commands
    which sequentially realize its move segments
    $\{\widetilde{m}\}$:
    $\{\epsilon_1,\ldots,\epsilon_k\}$;
    \\
    Update array
    $\mathbf{X}'_{\mathrm{No\text{-}Par}}
    =\epsilon_k\circ\cdots\circ\epsilon_1(\mathbf{X})$
};

\node (judge) [decision, below=11mm of seqstate, text width=54mm]
{
    $\mathbf{X}'_{\mathrm{Par}}
    =
    \mathbf{X}'_{\mathrm{No\text{-}Par}}$?
};

\node (append) [process, below=11mm of judge, text width=84mm]
{
    Append $m_j$ to $\widetilde{\epsilon}$;
    remove $m_j$ from $\mathcal{M}$
};

\node (store) [process, below=14mm of append]
{
    Store $\widetilde{\epsilon}$
    in parallel move list $\mathcal{S}$
};

\node (return) [io, below=14mm of store, text width=86mm]
{
    Return parallel move list
    $\mathcal{S}
    =
    \{\epsilon_1,\epsilon_2,\ldots,\epsilon_{M'}\}$
};

\node (end) [startstop, below=of return]
{
    Finish parallelization
};

% ------------------------------------------------
% Main downward arrows
% ------------------------------------------------

\draw[flowarrow] (start) -- (input);

\draw[flowarrow] (input) -- (nonempty);

\draw[flowarrow]
    (nonempty)
    --
    node[branchlabel, left] {True}
    (init);

\draw[flowarrow] (init) -- (iterate);

\draw[flowarrow]
    (iterate)
    --
    node[branchlabel, left] {Assign $i$}
    (parstate);

\draw[flowarrow] (parstate) -- (seqstate);

\draw[flowarrow] (seqstate) -- (judge);

\draw[flowarrow]
    (judge)
    --
    node[branchlabel, left] {True}
    (append);

%\draw[flowarrow] (store) -- (return);

\draw[flowarrow] (return) -- (end);

% ------------------------------------------------
% False branch from outer decision
% ------------------------------------------------

\coordinate (outerRight) at ($(nonempty.east)+(35mm,0)$);

\draw[flowarrow]
    (nonempty.east)
    --
    node[branchlabel, above] {False}
    (outerRight)
    |-
    (return.east);

% ------------------------------------------------
% End branch from iteration decision
% ------------------------------------------------

\coordinate (iterRight) at ($(iterate.east)+(18mm,0)$);

\draw[flowarrow]
    (iterate.east)
    --
    node[branchlabel, above] {End}
    (iterRight)
    |-
    (store.east);

% ------------------------------------------------
% Loop back to iterate from judge/append
% ------------------------------------------------

\coordinate (loopLeft) at ($(iterate.west)+(-18mm,0)$);
\coordinate (judgeLeft) at ($(loopLeft |- judge.west)$);
\coordinate (appendLeft) at ($(loopLeft |- append.west)$);

\draw[flowarrow]
    (judge.west)
    --
    node[branchlabel, above] {False}
    (judgeLeft)
    |-
    (iterate.west);

\draw[flowarrow]
    (append.west)
    --
    (appendLeft)
    |-
    (iterate.west);

% ------------------------------------------------
% Loop back to outer decision from store
% ------------------------------------------------

\coordinate (storeLeft) at ($(nonempty.west)+(-35mm,0)$);

\draw[flowarrow]
    (store.west)
    --
    (storeLeft |- store.west)
    |-
    (nonempty.west);

\end{tikzpicture}

\caption{Flowchart of the move parallelization subprocess.}
\label{fig:make_paralell}
\end{figure}

\paragraph{Move parallelization subroutine.}
\label{par:par_description}
% Parallelization of the atom moves is the key subprocess of ParHungarian, enhancing the rearrangement efficiency of Hungarian algorithm.
% In ParHungarian, the design of parallelizing moves process is based on the nature of AOD and follows the strategy of greedy algorithm.
Abstractly, given a sequence of individual atom moves $\mathcal{M}=\{m_1, \ldots, m_M\}$ and an atom array $\mathbf{X}$, we divide $\mathcal{M}$ into a set of `AOD moves' $\{\epsilon_1, \ldots, \epsilon_{M'}\}$.
% , where all moves in a subsequence $\Tilde{m}_i$ are executed in parallel.
Such parallelization is done following a greedy approach: the individual moves in $\mathcal{M}$ are iteratively added to a temporary parallel move queue $\Tilde{\epsilon}$, and subsequently eliminated from $\mathcal{M}$.
After each step, the algorithm checks whether $\Tilde{\epsilon}$ is a valid AOD move, and halts if not, removing the most recently added element. Then, a new list is created and the cycle repeats until all the moves in $\mathcal{M}$ have been sorted.
% Popping out the first item of $M$, i.e. $m_2$, ParHungarian inserts the item into $\Tilde{m}$.
% Now $\Tilde{m}$ is composed of $m_1$ and $m_2$.

To test if the move segments $\{\Tilde{m}_i\}$ in $\Tilde{\epsilon}$ can be implemented in parallel, the algorithm has to check whether parallelization of these moves affects the final configuration of the array.
If not, $\Tilde{\epsilon}$ is a valid AOD move and its composite move segments can be parallelized.
% In the experiment, we move atoms in the atom array by controlling and changing AODs' commands, so, abstractly, every move or parallel moves corresponds to an AOD command status, denoted as mapping $m \to \epsilon$.
% Following the aforementioned convention, the AOD command corresponding to execute $\Tilde{m}$ in parallel is expressed as $\Tilde{\epsilon}$ and 
To check whether this is the case, the composite AOD move $\Tilde{\epsilon}$ is applied to one copy of the atom array and another set – the AOD moves corresponding to executing each individual move segment sequentially (i.e. $\{\epsilon_1, \ldots, \epsilon_k\}$, where $k$ is defined as the number of move items in $\{\Tilde{m}\}$) – is applied to a second copy.
More formally, we compare the matrices $\mathbf{X}'_{\text{Par}}=\Tilde{\epsilon}(\mathbf{X})$ and $\mathbf{X}'_{\text{No-Par}}=\epsilon_k\circ\cdots\circ\epsilon_1(\mathbf{X})$.
If $\mathbf{X}'_{\text{Par}}=\mathbf{X}'_{\text{No-Par}}$, the moves $\{\Tilde{m}_i\}$ can be parallelized.

% If $m_2$ can be parallelized, the algorithm tries to add $m_3, \ldots, m_M$ into $\Tilde{m}$ and see if it can be parallelized with $m_1$ and $m_2$ step-by-step, and finally obtain a list of parallelizable moves $\Tilde{m}$.
% Extracting the moves in $\Tilde{m}$, the algorithm initializes $\Tilde{m}$ and inserts the first remaining item in $\mathcal{M}$ into $\Tilde{m}$.
 % Repeatedly executing the instructions in the last paragraph, the algorithm completes a new parallel move list $\{\Tilde{m}_1, \ldots, \Tilde{m}_{M'}\}$, where $m'$ is a single move or multiple parallel moves object and $M'$ is the size of new move list with the constraint $M'\leq M$.

\paragraph{Advantage and limitation of the parallelization strategy.}
Since ParHungarian checks whether any arbitrary set of moves can be parallelized, it potentially offers a higher degree of parallelism than algorithms which only parallelize moves within the space of a single row or column.
% Since ParHungarian determines if the moves can be parallelized or not by fundamental AOD commands, it offers more chance to parallelize the moves due to higher geometry flexibility compared with existing algorithms using fixed row/column parallelization strategy.

However, ParHungarian does not offer the optimal parallelization strategy due to its greedy approach.
The brute-force solution takes cost of $\mathcal{O}(2^M)$ to compute, where $M$ is the number of moves to be parallelized, since each move element can be parallelized and not parallelized.
To make the algorithm scalable, we adapt the greedy approach to handle the tradeoff of degree of parallelization and computational efficiency.
The computational complexity of our greedy parallelization strategy is $\mathcal{O}(M^2)$.

\subsection{Parallel Hungarian (dual-species)}
\label{sec:par_hung_dual}

\paragraph{Notations for dual-species atom arrays.}
The index set $\{1,\ldots,I\}$ is denoted by $[ I ]$, where $I\in\mathbb{N}$.
Here, we consider dual-species atom arrays $\mathbf{M}\in\mathbb{R}^{m\times n \times 2}$ as a stack of the $1^{\text{st}}$-species $\mathbf{M}_{1\text{st}}\in\mathbb{R}^{m\times n\times 1}$ and the $2^{\text{nd}}$-species $\mathbf{M}_{2\text{nd}}\in\mathbb{R}^{m\times n\times 1}$ arrays, where $m$ and $n$ denote number of rows and columns of the atom arrays and the third dimension denotes different atomic species/isotopes.
Due to the experimental constraints, the dual-species state matrix is subject to 
\begin{align}
\begin{cases}
    & \mathbf{M}_{ijk}\in\{0,1\} \quad\forall i \in [m], j\in[n], k\in[2]\\
    & \sum_{k=1}^2 \mathbf{M}_{ijk}\leq 1 \quad \forall i \in [m], j\in [n].
\end{cases}
\end{align}
The first constraint affirms that there are only two possible states for single sites, 0 (1) representing a vacancy (trapped atom).
% ``$0$''s standing for vacancies and ``$1$''s standing for trapped atoms.
The second constraint represents that a single site can only trap a single atom.
% Target atom arrays, $\mathbf{T}\in\mathbb{R}^{m\times n\times 2}$, is the desired configuration after we apply the InsideOut algorithm on inital dual-species atom arrays $\mathbf{M}$.
% Here, ``$1$''s denote sites expected to trap atoms and the third index denotes which species of atoms expected to trap.

\paragraph{Detailed description of dual-species ParHungarian.}
Additionally, we develop a naive extension of ParHungarian for dual-species rearrangement.
Fundamentally, the difference between the single-species and dual-species versions is the pairing and searching path processes.
The design of the algorithm closely follows the single-species version shown in Fig.~\ref{fig:par_Hung_flow} with two main differences:
\begin{itemize}
    \item
    \textbf{Matching atoms and targets}: The pairing process is species-selected.
    The algorithm needs to categorize the excess atoms and target vacancies as excess 1$^{\text{st}}$-species and 2$^{\text{nd}}$-species atoms and target vacancies.
    % for 1$^{\text{st}}$-species and 2$^{\text{nd}}$-species atoms.
    By repeating the cost matrix generation and matching process separately for each species, the dual-species ParHungarian obtains two sets of assignment pairs.
    \item
    \textbf{Searching path}: The searching path process has to take hetero-species obstacle atoms into consideration.
    In the single-species framework, even if there is no direct path between the source atom and the intended target, the domino segment approach is used as an alternative solution.
    However, in the dual-species framework, not all obstacles can be resolved with this strategy.
    % considered as the intermediate objects of domino moves.
    For example, if there is a $2^{\text{nd}}$-species atom located on the path between a $1^{\text{st}}$-species atom and a $1^{\text{st}}$-species vacancy, a `blocked' configuration is realized (see Fig.~\ref{fig5}a,b), and the rearrangement fails.
    % In our naive extension, ParHungarian only aims to search direct paths or homo-species segment paths (domino).
    % If these two types of path cannot be found, i.e. blocked cases illustrated in Fig.~\ref{fig5}a,b, the algorithm skips the path search process for this pair.
    % Therefore, sometimes dual-species fails to rearrange target patterns, as shown in Fig.~\ref{fig5}c,d.
\end{itemize}

\begin{figure}[p]
\centering

\begin{tikzpicture}[
    scale=0.60,
    transform shape,
    font=\footnotesize,
    node distance=5mm and 14mm,
    flowarrow/.style={
        -Latex,
        line width=0.75pt
    },
    startstop/.style={
        rectangle,
        rounded corners=2mm,
        draw,
        minimum height=9mm,
        inner xsep=4mm,
        align=center
    },
    io/.style={
        trapezium,
        trapezium left angle=70,
        trapezium right angle=110,
        draw,
        minimum height=10mm,
        text width=82mm,
        align=center,
        inner ysep=2mm
    },
    process/.style={
        rectangle,
        draw,
        minimum height=10mm,
        text width=78mm,
        align=center,
        inner sep=2.2mm
    },
    decision/.style={
        diamond,
        draw,
        aspect=3.6,
        text width=68mm,
        align=center,
        inner sep=1.3mm
    },
    branchlabel/.style={
        font=\footnotesize,
        fill=white,
        inner sep=1pt
    }
]

% ------------------------------------------------
% Main vertical sequence
% ------------------------------------------------

\node (start) [startstop]
{Start rearrangement};

\node (input) [io, below=of start]
{
    Initial loading array
    $\mathbf{M}_0\in\{0,1\}^{m\times n\times 2}$
    \\
    Target pattern
    $\mathbf{T}\in\{0,1\}^{m\times n\times 2}$
};

\node (initlayer) [process, below=of input]
{
    Initialize layer index $k=0$
};

\node (enough) [decision, below=8mm of initlayer]
{
    Are there sufficient atoms of both species,
    and is the rearrangement incomplete?
};

\node (parameterize) [process, below=9mm of enough]
{
    Increase the layer index, $k\leftarrow k+1$;
    \\
    parameterize the coordinates $(i,j)$ on this layer,
    denoted by $\mathcal{L}$.
    \\[-0.5mm]
    {\scriptsize
    \textsuperscript{a}See Eq.~\ref{eqn:k_layer_params}
    for the explicit parameterization rule.}
};

\node (layercomplete) [decision, below=9mm of parameterize]
{
    Is the rearrangement of the current layer incomplete?
    \\
    $\mathbf{M}_{ij1}=\mathbf{T}_{ij1}$ and
    $\mathbf{M}_{ij2}=\mathbf{T}_{ij2}$
    for all $(i,j)\in\mathcal{L}$
};

\node (push) [process, below=9mm of layercomplete]
{
    Push all misplaced atoms on the layer outward,
    edge by edge;
    \\
    update array $\mathbf{M}$.
    \\[-0.5mm]
    {\scriptsize
    An atom at $(i,j)$ is misplaced when its occupation
    does not match the target occupation.}
};

\node (assign) [process, below=of push]
{
    Generate $N$ species-selective atom--target pairs
    with minimum total distance.
    \\[-0.5mm]
    {\scriptsize
    \textsuperscript{b}Targets are empty sites on the layer.
    \quad
    \textsuperscript{c}Atoms are selected from on or outside
    the layer.}
};

\node (paths) [process, below=of assign]
{
    Find the shortest path for each pair,
    denoted by $\{p_\mu\}_{\mu=1}^{N}$.
    \\
    Let
    $p_\mu:=\{m_{\mu t}\}_{t=1}^{L_\mu}$
    and
    $L:=\max_\mu\operatorname{Len}(p_\mu)$.
};

\node (sorted) [decision, below=8mm of paths]
{
    Have all moves in every path been sorted?
};

\node (collect) [process, below=9mm of sorted]
{
    Remove the first element from each nonempty path
    and add the elements to a new list $q_\mu$.
    \\[-0.5mm]
    {\scriptsize
    \textsuperscript{d}A path may be empty because
    path lengths can differ.}
};

\node (parallel) [process, below=of collect]
{
    Check for collisions among the moves in $q_\mu$.
    If a collision occurs, retain one move in $q_\mu$
    and return the others to their corresponding paths $p_\mu$.
    \\
    Parallelize the retained moves
    (Fig.~\ref{fig:make_paralell}).
};

\node (execute) [
    process,
    text width=94mm,
    below=of parallel
]
{
    Construct a sequence of operations
    $\widetilde{\epsilon}
    =\{\epsilon_\nu\}_{\nu=1}^{N'}$,
    where $N'\leq N$.
    \\
    Execute the moves and update the array:
    $\mathbf{M}_{t+N'}
    =\widetilde{\epsilon}(\mathbf{M}_t)$.
};

\node (return) [
    io,
    text width=60mm,
    below=7mm of execute
]
{
    Return final array
    $\mathbf{M}_{\mathrm{final}}$
};

\node (end) [startstop, below=of return]
{
    Finish rearrangement
};

% ------------------------------------------------
% Main downward arrows
% ------------------------------------------------

\draw[flowarrow] (start) -- (input);
\draw[flowarrow] (input) -- (initlayer);
\draw[flowarrow] (initlayer) -- (enough);

\draw[flowarrow]
    (enough)
    --
    node[branchlabel, left] {True}
    (parameterize);

\draw[flowarrow] (parameterize) -- (layercomplete);

\draw[flowarrow]
    (layercomplete)
    --
    node[branchlabel, left] {True}
    (push);

\draw[flowarrow] (push) -- (assign);
\draw[flowarrow] (assign) -- (paths);
\draw[flowarrow] (paths) -- (sorted);

\draw[flowarrow]
    (sorted)
    --
    node[branchlabel, left] {Assign moves}
    (collect);

\draw[flowarrow] (collect) -- (parallel);
\draw[flowarrow] (parallel) -- (execute);
%\draw[flowarrow] (execute) -- (return);
\draw[flowarrow] (return) -- (end);

% ------------------------------------------------
% Layer-complete branch:
% return to the global stopping condition
% ------------------------------------------------

\coordinate (layerLeft)
    at ($(layercomplete.west)+(-22mm,0)$);

\draw[flowarrow]
    (layercomplete.west)
    --
    node[branchlabel, above] {False}
    (layerLeft)
    |-
    (enough.west);

% ------------------------------------------------
% Global stopping branch:
% terminate and return the final array
% ------------------------------------------------

\coordinate (outerRight)
    at ($(enough.east)+(32mm,0)$);

\draw[flowarrow]
    (enough.east)
    --
    node[branchlabel, above] {False}
    (outerRight)
    |-
    (return.east);

% ------------------------------------------------
% End of current layer:
% return to layer-completion decision
% ------------------------------------------------

\coordinate (layerRight)
    at ($(sorted.east)+(20mm,0)$);

\draw[flowarrow]
    (sorted.east)
    --
    node[branchlabel, above] {End}
    (layerRight)
    |-
    (layercomplete.east);

% ------------------------------------------------
% Continue sorting moves in the current layer
% ------------------------------------------------

\coordinate (sortLeft)
    at ($(execute.west)+(-20mm,0)$);

\draw[flowarrow]
    (execute.west)
    --
    (sortLeft)
    |-
    (sorted.west);

\end{tikzpicture}

\captionsetup{skip=3pt}
\caption{Flowchart of the InsideOut algorithm.}
\label{fig:in_out_flow}
\end{figure}

\subsection{InsideOut Algorithm (dual-species)}
\label{sec:inside_out}
To overcome the unique challenges of dual-species rearrangement (discussed in Sec.~\ref{sec:dual}), we develop the InsideOut algorithm.
The cornerstone of InsideOut is its simple and uniform layer-by-layer approach, which effectively prevents the emergence of `blocked cases' shown in Fig.~\ref{fig5}a,b.
In case blocked configurations still occur, InsideOut pushes the obstacle atoms away from the center in a `clearing' step.
% still has cleaning path function that pushes out obstacles if there is no available path between atoms and target vacancies.
Visualization of the algorithm is shown in Fig.~\ref{fig:in_out_flow}.

\paragraph{Detailed description of InsideOut.}
First, InsideOut checks if the number of atoms in both species is sufficient to prepare the target pattern, i.e. 
\begin{align}
\begin{cases}
    & \sum_{ij} \mathbf{M}_{ij1} \geq \sum_{ij} \mathbf{T}_{ij1}\\
    &\sum_{ij} \mathbf{M}_{ij2} \geq \sum_{ij} \mathbf{T}_{ij2}.
\end{cases}
\end{align}
If not, the algorithm halts.

Then, InsideOut parameterizes the first layer of the atom array and aims to rearrange all sites on this layer.
Given layer index $k$ ($k=1$ for the first layer), the coordinates on the $k$th layer can be expressed as
\begin{equation}
\label{eqn:k_layer_params}
\begin{aligned}
\mathcal{L}_k =\;&
\{\,(\mathrm{top},\,j')\mid j'=\mathrm{left},\dots,\mathrm{right}\}\\
&\cup\;\{\, (i',\,\mathrm{right})\mid i'=\mathrm{top}+1,\dots,\mathrm{bottom}\}\\
&\cup\;\{\,(\mathrm{bottom},\,j')\mid j'=\mathrm{right}-1,\dots,\mathrm{left}\}\\
&\cup\;\{\, (i',\,\mathrm{left})\mid i'=\mathrm{bottom}-1,\dots,\mathrm{top}+1\}.
\end{aligned}
\end{equation}
with the parameters
\begin{equation}
\begin{aligned}
\delta_\text{row} &\equiv (m+1)\bmod 2,
&\quad
\delta_\text{col} &\equiv (n+1)\bmod 2,\\
c_\text{row} &= \Bigl\lfloor \tfrac{m}{2} \Bigr\rfloor - \delta_\text{row} + 1,&\quad
c_\text{col} &= \Bigl\lfloor \tfrac{n}{2} \Bigr\rfloor - \delta_\text{col} + 1,\\
\mathrm{top}    &= c_\text{row} - (k-1), 
&\quad
\mathrm{left}   &= c_\text{col} - (k-1),\\
\mathrm{bottom} &= c_\text{row} + (k-1) + \delta_\text{row},
&\quad
\mathrm{right}  &= c_\text{col} + (k-1) + \delta_\text{col},
\end{aligned}
\end{equation}
where $\delta_\text{row}$ and $\delta_\text{col}$ are factors used to handle odd/even number of rows/columns, $c_\text{row}$ and $c_\text{col}$ are the row/column coordinates of the center of the target array, $\mathrm{top}/\mathrm{left}/\mathrm{bottom}/\mathrm{right}$ are the boundaries of the $k$th layer.

After identifying sites on the $k$th layer, InsideOut pushes out misplaced atoms on the $k$th layer, i.e. filled sites with coordinate of $(i, j)$ satisfying both conditions of $\mathbf{M}_{ij1}\neq\mathbf{T}_{ij1}$ and $\mathbf{M}_{ij2}\neq\mathbf{T}_{ij2}$ to the outer layer.
Note that the direction in which obstacles are pushed is determined by which edge the atom is located at in the expression of Eq.~\ref{eqn:k_layer_params}.
For an obstacle on the top edge, we push the atom upwards.
Furthermore, to accelerate the `clearing' process, all misplaced atoms on the same edge are pushed away in parallel.

Similar to the ParHungarian algorithm, InsideOut computes two cost matrices $\mathbf{C}_{1\text{st}}$ and $\mathbf{C}_{2\text{nd}}$ corresponding to two species of atoms with the element of Euclidean distance between targets and atoms.
By solving two linear sum assignment problems corresponding to  $\mathbf{C}_{1\text{st}}$ and $\mathbf{C}_{2\text{nd}}$ independently, InsideOut generates $N'$ atom-target pairs.
Except for the dual-species searching path constraints discussed in \ref{sec:par_hung_dual}, the searching path and moving atoms process of InsideOut follows those of ParHungarian.

After all targets on the $k$th layer are correctly filled, InsideOut moves to the next layer ($k+1$). This process is repeated, rearranging atoms layer-by-layer, until the target pattern $\mathbf{T}$ is realized.

\end{document}